\documentclass[5p,times,numbers,sort&compress]{elsarticle}  % natbib handled internally\usepackage{url}
\usepackage{graphicx}
\usepackage{booktabs}
\usepackage{amsmath,amssymb}
\usepackage{newtxtext,newtxmath} % Times风格文本+改进的数学字体
\newtheorem{assumption}{Assumption}
\usepackage{pgfgantt}
\usepackage{hyperref}
\usepackage{mathrsfs}
\usepackage{subcaption}
\usepackage{textcomp}
\usepackage{xcolor}
\usepackage{enumitem}
\usepackage{multirow}
\newproof{pf}{Proof}

\usepackage{algorithm}        % 提供算法计数器（我们不会用它的浮动体）
\usepackage{algpseudocode}    % algorithmicx 的伪代码环境
\usepackage[most]{tcolorbox}  % 可跨页盒子（breakable）
\usepackage{caption}          % 备用的 captionof 等，不必须但推荐
\algrenewcommand\algorithmicrequire{\textbf{Inputs:}}
\algrenewcommand\algorithmicensure{\textbf{Output:}}
\usepackage{stfloats}

\journal{}   % or Energy

\begin{document}

\begin{frontmatter}

\title{Scalable Distributed Learning-Based MPC for Heterogeneous Consumer--Prosumer Building Aggregations: MPC-Aware Feature Selection and Convex Constraint-Coupled Optimization}

\author[1]{Kaipeng Xu}
\author[1]{Zhuo Zhi}
\author[1]{Keyue Jiang}

\address[1]{Department of Electronic and Electrical Engineering, University College London, Torrington Place, London WC1E 7JE, United Kingdom}

% Optional abstract placeholder (no content change)
\begin{abstract}
    Buildings provide demand-response flexibility through controllable thermal loads and thermal inertia, while distributed photovoltaic deployment creates opportunities to share surplus generation across heterogeneous consumer--prosumer aggregations. Exploiting this flexibility at aggregation scale requires scalable real-time coordination.  This study develops a scalable distributed learning-based model predictive control (MPC) framework that preserves convexity to enable a fully distributed optimizer with an asymptotic global-optimality guarantee. The previously developed Input-Convex Encoder-Only Transformer (IC-EoT) is embedded through recursive substitution to preserve local MPC convexity, while the aggregation-wide energy-trading balance yields a convex constraint-coupled problem solved by Tracking Alternating Direction Method of Multipliers (Tracking-ADMM) with bounded-neighbour communication and fully distributed finite-iteration stopping. An MPC-aware feature-selection procedure additionally accounts for future-value availability and recursive error propagation. In EnergyPlus--Energym co-simulation, centralized MPC becomes unable to meet the real-time solution requirement when the aggregation size reaches six buildings. In contrast, distributed MPC (DMPC) remains real-time feasible throughout the tested range and shows only mild solution-time growth, accompanied primarily by increasing Tracking-ADMM iterations as aggregation size grows. Where centralized references are available, finite-iteration DMPC approaches the centralized objective while satisfying the stopping tolerances. Coordination yields bill savings of up to 9.6\% when sufficient photovoltaic surplus and consumer demand coincide, whereas low-surplus savings can be masked by closed-loop numerical variation associated with finite solver convergence tolerances. The proposed feature sets reduce horizon-mean weighted standardized primary-target mean squared error by up to 9.7\% and 33.8\% for the office and apartment testbeds, respectively.
\end{abstract}


\end{frontmatter}

\section{Introduction}
\label{sec:introduction}

Buildings are among the largest global energy consumers, accounting for approximately 32\% of global energy demand and 34\% of global CO$_2$ emissions \cite{unep2025buildings}. Heating, Ventilation, and Air Conditioning (HVAC) systems constitute a substantial share of building energy use \cite{drgovna2020all}. The substantial energy footprint and inherent thermal storage capacity of buildings present a prime opportunity for Demand Response (DR) programs. By acting as a "virtual battery," buildings can pre-heat or pre-cool during off-peak hours, shifting energy consumption away from high-price periods to reduce costs and support grid stability \cite{ma2012demand, vedullapalli2019combined}.

Model Predictive Control (MPC) is widely used for HVAC-based DR because it can explicitly handle multivariable constraints while optimizing control actions over a future prediction horizon \cite{drgovna2020all}. The efficacy of MPC, however, is critically dependent on its internal predictive model. While first-principles models are time-consuming to develop \cite{moriyasu2021structured}, data-driven models using machine learning offer a more flexible and often more accurate alternative for capturing complex whole-building thermal and energy consumption dynamics \cite{gonzalez2017data, azuatalam2020reinforcement}.

The predictive performance of data-driven models is strongly influenced by the choice of input features. Modern building automation systems can provide large numbers of candidate variables, making systematic feature selection important for reducing dimensionality and improving model generalization \cite{zhang2019systematic}. Existing building-energy feature-selection procedures provide effective combinations of domain knowledge, statistical filtering, and wrapper evaluation \cite{zhang2019systematic}. However, conventional forecasting-oriented selection does not explicitly account for how each selected variable will be deployed during recursive MPC prediction. In particular, a variable whose future value is unknown must itself be predicted once the model is rolled forward, so its prediction error can propagate through the remainder of the MPC horizon. Consequently, a feature that improves one-step prediction does not necessarily improve the multi-step predictions used by the controller \cite{taieb2012review}. This motivates a feature-selection procedure that evaluates candidate variables according to their actual future availability and recursive prediction roles in MPC.

The transition from individual buildings to coordinated building aggregations creates an additional opportunity. Increasing deployment of distributed photovoltaic (PV) enables locally generated electricity to be shared among participants rather than being treated independently at each building, allowing surplus generation from one participant to offset demand elsewhere in the aggregation. Such collective operation is becoming increasingly relevant in practice: the European Commission's Distribution System Operator (DSO) Observatory, covering operators across 26 European Union Member States, reports more than 247~GW of renewable capacity connected to the surveyed networks, of which 52\% is PV, while around 12,400 energy communities were reported in the networks of half of the participating operators \cite{meletiou2025dso}. The same report highlights the introduction of a regulatory ``right to energy sharing'', reinforcing the practical relevance of coordinated local energy exchange. Local energy markets and collective self-consumption therefore provide a practical context in which coordinated control of multiple buildings can exploit energy exchanges that are unavailable when each building is operated independently \cite{capper2022localmarkets}.

Such aggregations need not be homogeneous in their distributed energy resource (DER) ownership or market roles. In this study, a \textbf{prosumer} denotes a building equipped with on-site PV generation and capable of exporting surplus electricity to the main grid or trading it within the aggregation, whereas a \textbf{consumer} has no on-site generation and purchases electricity from either the main grid or the internal aggregation market. Both roles retain demand-side flexibility through HVAC control and building thermal inertia, while consumer buildings can additionally serve as flexible recipients of locally available surplus generation. Moreover, the transition from conventional consumers to prosumers remains gradual rather than universal, making mixed consumer--prosumer populations a practically relevant aggregation structure \cite{grzanic2022prosumers}. Coordinating such heterogeneous agents creates the opportunity to match locally available PV surplus with flexible demand elsewhere in the aggregation, while coupling otherwise independent building-level control decisions through the internal energy balance.

Coordinating these buildings through a single centralized MPC, however, introduces a scalability challenge because the joint decision space and constraint set grow with the aggregation size \cite{diekerhof2017hierarchical, li2021parallel}. This computational growth can ultimately conflict with the fixed real-time solution interval required by MPC. Distributed optimization provides a natural alternative by retaining building-level decision variables locally while coordinating only the quantities required to satisfy aggregation-wide constraints.

The aggregation-wide energy balance gives the resulting MPC the structure of a constraint-coupled problem (CCP). For this problem, the aim is to obtain a fully distributed solution without a central coordinating unit while retaining an asymptotic global-optimality guarantee. Tracking Alternating Direction Method of Multipliers (Tracking-ADMM) provides these properties for convex local problems subject to a common linear coupling constraint \cite{falsone2020tracking}. Distributed methods for certain non-convex problems also exist, but their guarantees are generally limited to stationary solutions under additional problem-specific conditions \cite{hong2016nonconvex}. Preserving convexity in the local learning-based MPC problems is therefore a deliberate modelling choice of the proposed framework rather than a generic requirement of distributed optimization.

Input-Convex Neural Networks (ICNNs) provide a structural route to satisfying this requirement by constraining the network architecture so that its output is convex with respect to selected inputs, thereby enabling those inputs to be optimized within a convex formulation \cite{amos2017input}. Temporal input-convex architectures extend the ICNN paradigm to sequential dynamical modelling, including the Input-Convex Recurrent Neural Network (ICRNN) \cite{chen2018optimal} and the Input-Convex Long Short-Term Memory network (IC-LSTM) \cite{wang2025real}. For the temporal building dynamics considered here, the previously developed Input-Convex Encoder-Only Transformer (IC-EoT) \cite{xu2026input} is adopted as the predictive model. IC-EoT provides an attention-based alternative for sequence modelling while preserving the input-convex structure required by the subsequent convex MPC formulation. It was shown to achieve substantially lower MPC solution times than IC-LSTM while maintaining comparable predictive and closed-loop control performance \cite{xu2026input}, making it particularly suitable for the repeated local optimizations required by distributed MPC.

These challenges motivate an integrated methodology in which feature selection is aligned with recursive MPC deployment, learning-based prediction preserves the convexity required by the selected distributed optimizer, and aggregation-wide coordination remains computationally viable for real-time control. The main contributions are threefold:

\begin{itemize}[leftmargin=*]
    \item \textbf{Convex learning-based distributed MPC for heterogeneous aggregation:} The previously developed IC-EoT \cite{xu2026input} is integrated into consumer and prosumer MPC sub-problems through a recursive formulation that preserves convexity with respect to the local control decisions. Together with the affine aggregation-wide energy-trading balance, the resulting control problem is formulated as a convex constraint-coupled problem compatible with the established convergence conditions of Tracking-ADMM \cite{falsone2020tracking}.
     
    \item \textbf{Scalable fully distributed MPC implementation and validation:} Tracking-ADMM is implemented for receding-horizon building coordination using bounded-neighbour communication and fully distributed finite-iteration stopping. The framework is evaluated against Individual MPC and centralized MPC (CMPC) through complementary scalability, convergence, and coordination studies. The results show that centralized solution time increases rapidly with aggregation size, whereas distributed MPC (DMPC) exhibits substantially milder computational growth while approaching the centralized benchmark; the coordination experiments further show that observable economic benefits depend on sufficient redistributable PV surplus and consumer demand capable of absorbing it.
    
    \item \textbf{MPC-aware feature selection for recursive prediction:} Building on the systematic three-stage framework of \cite{zhang2019systematic}, the proposed procedure evaluates candidate variables according to their future availability and recursive deployment roles in MPC. Candidate features whose future values are unavailable are incorporated into the recursive prediction set during wrapper evaluation, so that their propagated errors are assessed directly over the control-relevant prediction horizon. The resulting feature sets achieved the lowest horizon-mean and end-of-horizon primary-target errors across both building testbeds.
\end{itemize}

The remainder of this paper is organized as follows. Section~\ref{sec:related_work} reviews input-convex modelling for tractable MPC, feature selection for learning-based MPC, and distributed optimization for constraint-coupled problems. Section~\ref{sec:methodology} formulates the heterogeneous building-aggregation MPC as a convex constraint-coupled problem and presents the Tracking-ADMM solution framework and its convergence properties. Section~\ref{sec:feature_selection} develops the MPC-aware feature-selection methodology. Section~\ref{sec:case_study} presents the co-simulation case study, the finite-iteration distributed MPC implementation, and the experimental evaluation of feature selection, computational scalability, convergence, and coordination value. Finally, Section~\ref{sec:conclusion} summarizes the main findings and outlines directions for future work.

\section{Related Work}
\label{sec:related_work}

\subsection{Input-Convex Models for Tractable MPC}

The integration of standard neural-network dynamics models into MPC can result in non-convex optimization problems, complicating reliable real-time solution and removing global-optimality guarantees \cite{chen2018optimal,bunning2021input}. ICNNs, introduced by \cite{amos2017input}, impose architectural constraints such that the network output is convex with respect to designated inputs. This property enables efficient global optimization over those inputs, but the original ICNN formulation does not by itself address the recursive multi-step prediction required in MPC.

For dynamical systems, Chen et al.~\cite{chen2018optimal} extended the input-convex principle to ICRNNs, establishing convexity with respect to the control-input sequence and demonstrating the resulting convex MPC formulation in applications including building HVAC control. Feed-forward ICNNs were subsequently adapted specifically for recursive building MPC by Bünning et al.~\cite{bunning2021input}, who introduced additional structural restrictions to preserve convexity across the prediction horizon and demonstrated the approach in a real apartment. Input-convex models have also been applied to nonlinear MPC of power-generation systems \cite{zhu2022nonlinear} and optimization of chiller--pump systems \cite{xing2024bilevel}. More recently, IC-LSTM was proposed to provide richer temporal modelling while retaining input convexity for optimization \cite{wang2025real}.

Recurrent input-convex architectures nevertheless inherit the sequential computation of recurrent models, which can become restrictive when repeatedly evaluated within MPC optimization. The IC-EoT developed in the preceding work replaces recurrence with an encoder-only self-attention architecture while preserving the recursive input-convexity required by MPC \cite{xu2026input}. Its more favourable online MPC solution-time scaling relative to IC-LSTM motivates its use here, where the resulting convex local MPC problems are subsequently coordinated through distributed optimization.

\subsection{Feature Selection for Learning-Based MPC}

Beyond predictive-model architecture, the selection of informative and deployable input variables is also important for learning-based MPC. Building automation systems can expose large numbers of candidate measurements, making systematic feature selection valuable for reducing model complexity and improving generalization. Zhang and Wen~\cite{zhang2019systematic} proposed a systematic three-step procedure combining domain-knowledge-based pre-processing, filter-based removal of irrelevant and redundant variables, and a forward wrapper for final feature selection. Importantly for control applications, their procedure identifies required control variables as mandatory features and exempts them from subsequent removal.

More recent studies have brought feature selection closer to predictive-control requirements. Zhang~\cite{zhang2021data} integrated feature selection and active learning into the development of whole-building models for data predictive control. Zhao et al.~\cite{zhao2023feature} combined filter, wrapper, and embedded feature-selection methods with reduced-order recurrent neural network models for distributed MPC of nonlinear processes. For building MPC, Erfani et al.~\cite{erfani2024search} explicitly compared one-step and multi-step prediction errors when selecting predictive models and showed that multi-step performance provides a more informative indicator for flexibility activation. Shapley additive explanations (SHAP)-based feature ranking has also been incorporated into data-driven building MPC~\cite{henkel2024interpretable}, while subsequent work has further shown that variables important for predictive-model accuracy are not necessarily equally important for MPC performance~\cite{erfani2025does}.

These studies establish that feature selection for predictive control should account for the downstream control task rather than one-step model accuracy alone. In the studies reviewed above, however, the deployment role of an auxiliary feature selected for a recursively evaluated one-step model is not explicitly incorporated into the selection procedure. In particular, horizon-known candidates can remain exogenous inputs over the MPC horizon, whereas future-unknown candidates must themselves be recursively supplied if retained. Their prediction errors can consequently propagate through the rollout and affect the primary MPC predictions. The methodology developed in Section~\ref{sec:feature_selection} addresses this issue by distinguishing candidates according to future-value availability and evaluating the resulting input and recursive-target configuration through a free-running rollout over the MPC prediction horizon.

\subsection{Distributed Optimization for Constraint-Coupled Problems}

The building-aggregation problem considered in this study belongs to the class of constraint-coupled optimization problems, in which agents retain private local decision variables and objectives while their decisions are linked through a common coupling constraint. This structure differs from consensus or decision-coupled optimization, where agents seek agreement on copies of a common decision vector. Although algorithms developed for the latter class can in principle be reformulated for constraint-coupled problems, doing so may require each agent to maintain estimates of the complete global decision vector and to access the full coupling constraint~\cite{falsone2020tracking}. Dedicated constraint-coupled methods therefore better preserve the local decision structure required for scalable multi-agent control.

A number of constraint-coupled approaches are based on dual decomposition, primal--dual methods, or the alternating direction method of multipliers (ADMM). A classical parallel ADMM formulation decomposes the primal optimization across agents, while the coupling residual and common dual variable are coordinated centrally \cite{bertsekas1989parallel,falsone2020tracking}. Tracking-ADMM~\cite{falsone2020tracking} removes this central coordinator by embedding a dynamic average tracking mechanism into the parallel ADMM structure. Each agent consequently maintains local estimates of the network-wide coupling violation and dual variable through neighbour communication. For convex local objectives, convex compact local feasible sets, and affine equality coupling, the method provides asymptotic primal optimality and dual consensus without requiring strict or strong convexity.

Subsequent work has extended distributed optimization for constraint-coupled problems. Augmented Lagrangian Tracking (ALT)~\cite{falsone2023augmented} accommodates both affine equality and nonlinear inequality coupling constraints and removes the compactness requirement on local feasible sets. Importantly, when inequality coupling is absent, ALT reduces to Tracking-ADMM. More recently, Yarmoshik et al.~\cite{yarmoshik2025decentralized} developed a linearly convergent decentralized method for affine constraint-coupled problems under smooth and strongly convex local objectives. These assumptions are stronger than those imposed on the local MPC problems considered here, which are convex but are not required to be strongly convex.

Tracking-ADMM is therefore adopted because its assumptions directly match the present MPC formulation: the local problems are convex but not required to be strongly convex, and the aggregation is coupled through a common affine equality constraint. It also removes the central coordination required by parallel ADMM while providing each agent with a local estimate of coupling infeasibility, a property particularly useful for distributed receding-horizon control \cite{falsone2020tracking}.

\section{Constraint-Coupled MPC Formulation and Distributed Solution}
\label{sec:methodology}

This section develops the distributed MPC framework from the structure of the aggregation problem. First, the finite-horizon coordination problem is formulated as a CCP with an affine coupling constraint. This structure matches the problem class addressed by Tracking-ADMM \cite{falsone2020tracking}, which considers convex local objective functions, convex compact local constraint sets, and a common linear coupling constraint. Convexity is also valuable independently in learning-based MPC because it permits globally tractable optimization over the control inputs, an advantage that motivates input-convex models for optimization and MPC \cite{amos2017input,chen2018optimal,bunning2021input}. Among temporal input-convex models, IC-EoT is adopted because it combines nonlinear, non-recurrent sequence modelling with the recursive input-convexity required by MPC and has demonstrated favourable online MPC solution-time scaling relative to recurrent input-convex models \cite{xu2026input}. Tracking-ADMM is then applied to the resulting convex CCP, while the practical receding-horizon implementation is specified in Section~\ref{sec:case_study}.

\subsection{General Constraint-Coupled Formulation}
The aggregation comprises $N$ agents. The agent index set is $\mathcal{B}=\{1,\ldots,N\}$. At each MPC update time $t$, the complete finite-horizon decision vector of agent $i$ is denoted by $\boldsymbol{\xi}_t^i$. Depending on the agent type, this vector collects the HVAC control sequence together with the local energy-allocation and other auxiliary decision variables specified later in Section~\ref{sec:case_study}. The aggregation-level MPC problem is written in the canonical constraint-coupled form

\begin{subequations}\label{eq:general_ccp}
\begin{align}
\min_{\{\boldsymbol{\xi}_t^i\}_{i\in\mathcal B}} \quad & \sum_{i\in\mathcal B} J_t^i(\boldsymbol{\xi}_t^i) \label{eq:ccp_objective}\\
\text{s.t.}\quad & \boldsymbol{\xi}_t^i\in\mathcal F_t^i,\qquad i\in\mathcal B, \label{eq:ccp_local}\\
& \sum_{i\in\mathcal B}A_i\boldsymbol{\xi}_t^i=\mathbf b_t. \label{eq:ccp_coupling}
\end{align}
\end{subequations}

Here, $J_t^i$ is the local finite-horizon objective, $\mathcal F_t^i$ is the local feasible decision set, and $A_i\boldsymbol{\xi}_t^i$ is the contribution of agent $i$ to the affine aggregation-wide coupling constraint. The separable objective and common affine coupling constraint in \eqref{eq:general_ccp} place the aggregation MPC in the class of constraint-coupled problems considered by Tracking-ADMM \cite{falsone2020tracking}. Tracking-ADMM removes the central coordination required by the corresponding parallel ADMM formulation by locally tracking the coupling-constraint violation through neighbour communication. Its convergence and optimality results require convex local objectives and convex compact local feasible sets. Consequently, the predictive MPC formulation of each agent must preserve convexity with respect to its decision variables; the IC-EoT construction used to obtain this property is described next.

\subsection{Convex Local MPC Formulation via Input-Convex Temporal Modelling}

To satisfy the convexity requirement identified above without reverting to a linear building predictor, the IC-EoT developed in \cite{xu2026input} is employed. The model was designed specifically for learning-based MPC to retain nonlinear temporal modelling while preserving the input-convex structure required for convex optimization. In addition to providing a global-optimality interpretation for converged MPC solutions, this structure avoids the generic non-convex optimization induced by conventional neural predictors \cite{xu2026input}. For recursive multi-step MPC, convexity must also be preserved when predicted variables are substituted into subsequent prediction steps; the componentwise convex and non-decreasing structure established for IC-EoT provides this composition property under the input construction described below.

The IC-EoT predictor is denoted by $f_\theta$, where $\theta$ collects its trained model parameters. For each agent $i\in\mathcal B$, define the finite-horizon HVAC control sequence as
\begin{equation}
\mathbf U_t^i:=\operatorname{col}\!\left(\mathbf u_{t|t}^i,\mathbf u_{t+1|t}^i,\ldots,\mathbf u_{t+N_p-1|t}^i\right),
\label{eq:control_horizon}
\end{equation}
where $\mathbf u_{t+k|t}^i$ denotes the HVAC control applied at prediction stage $k\in\mathcal K:=\{0,\ldots,N_p-1\}$.

Let $\mathbf H_{0|t}^i$ denote the measured historical input window available to agent $i$ at MPC update time $t$, and let $\hat{\mathbf y}_{0|t}^i$ denote the corresponding current measured model-variable vector. At prediction stage $k$, the recursively predicted model variables and the candidate control input form the control-augmented row
\begin{equation}
\tilde{\boldsymbol{\zeta}}_{k|t}^i=
\begin{bmatrix}
(\hat{\mathbf y}_{k|t}^i)^\top &
(\mathbf u_{t+k|t}^i)^\top &
-(\mathbf u_{t+k|t}^i)^\top
\end{bmatrix}^{\!\top}.
\label{eq:selective_extension}
\end{equation}
Only the control variables are duplicated with opposite sign; the measured and recursively predicted model variables are not negatively duplicated, consistently with the recursive IC-EoT MPC formulation in \cite{xu2026input}. Any selected horizon-known auxiliary feature is appended to the corresponding input row as fixed problem data and is omitted from \eqref{eq:selective_extension} for notational compactness.

The fixed-length prediction history and the corresponding one-step prediction are propagated recursively as
\begin{align}
\mathbf H_{k+1|t}^i&=\operatorname{shift}\!\left(\mathbf H_{k|t}^i,\tilde{\boldsymbol{\zeta}}_{k|t}^i\right), \label{eq:history_update}\\
\hat{\mathbf y}_{k+1|t}^i&=f_\theta\!\left(\mathbf H_{k+1|t}^i\right),\qquad k\in\mathcal K, \label{eq:recursive_prediction}
\end{align}
where $\operatorname{shift}(\cdot)$ removes the oldest row of the history window and appends $\tilde{\boldsymbol{\zeta}}_{k|t}^i$. Hence, $\mathbf H_{k+1|t}^i$ is the complete IC-EoT input at prediction stage $k+1$, while $\hat{\mathbf y}_{k+1|t}^i$ is the corresponding model output. This output enters \eqref{eq:selective_extension} at the subsequent prediction stage, generating the recursive multi-step rollout over the MPC horizon.

By Theorem~1 and Corollary~2 of \cite{xu2026input}, the selective control extension in \eqref{eq:selective_extension} together with the recursive history propagation in \eqref{eq:history_update} preserves the required convex dependence of the predicted quantities on the complete control sequence $\mathbf U_t^i$ for fixed measured history $\mathbf H_{0|t}^i$. After recursively substituting the intermediate predictions, the predicted quantities entering the MPC problem can therefore be expressed as convex functions of $\mathbf U_t^i$ without introducing the neural predictor as a non-affine equality constraint between independent optimization variables. Consequently, the local MPC problem remains convex when these predictions enter through convex objective terms and convex sublevel constraints, while equality constraints involving optimization variables remain affine. The specific consumer and prosumer formulations are given in Section~\ref{sec:case_study}. For the distributed optimization analysis, the resulting local MPC problems are summarized through the convexity and compactness conditions required by Tracking-ADMM:

\begin{assumption}[Convexity and compactness]
For each agent $i\in\mathcal B$, $J_t^i$ is convex and $\mathcal F_t^i$ is non-empty, convex, and compact. The aggregation-wide coupling constraint \eqref{eq:ccp_coupling} is affine.
\label{ass:convex_ccp}
\end{assumption}

\subsection{Distributed Solution of the Convex Constraint-Coupled Problem}

Under Assumption~\ref{ass:convex_ccp}, Tracking-ADMM \cite{falsone2020tracking} can be applied to \eqref{eq:general_ccp}. The MPC-time index $t$ is fixed during each distributed solve and is therefore suppressed below, while $\tau$ denotes the Tracking-ADMM iteration index. The local tracking variable and local dual estimate are denoted by $\boldsymbol{\psi}_i^\tau$ and $\boldsymbol{\nu}_i^\tau$, respectively.

Let $\mathcal N_i$ denote the neighbour set of agent $i$, including agent $i$ itself.

\begin{assumption}[Tracking-ADMM regularity conditions]
The CCP in \eqref{eq:general_ccp} and its Lagrangian dual admit optimal solutions. The communication graph is fixed, undirected, and connected. The mixing matrix $\mathbf M=[m_{ij}]$ is symmetric, doubly stochastic, positive semidefinite, and satisfies $m_{ij}>0$ if and only if $j\in\mathcal N_i$.
\label{ass:tracking_admm}
\end{assumption}

Let $\mathbf b_i$ satisfy $\sum_{i\in\mathcal B}\mathbf b_i=\mathbf b_t$. Given an initial feasible local decision $\boldsymbol{\xi}_i^0\in\mathcal F_i$, an initial dual estimate $\boldsymbol{\nu}_i^0$, and the tracker initialization $\boldsymbol{\psi}_i^0=A_i\boldsymbol{\xi}_i^0-\mathbf b_i$, the Tracking-ADMM iteration of \cite{falsone2020tracking} is expressed in the notation of \eqref{eq:general_ccp} as
\begin{subequations}\label{eq:tracking_admm}
\begin{align}
\bar{\boldsymbol{\nu}}_i^\tau&=\sum_{j\in\mathcal N_i}m_{ij}\boldsymbol{\nu}_j^\tau,\qquad \bar{\boldsymbol{\psi}}_i^\tau=\sum_{j\in\mathcal N_i}m_{ij}\boldsymbol{\psi}_j^\tau, \label{eq:tracking_mix}\\
\boldsymbol{\xi}_i^{\tau+1}&\in\underset{\boldsymbol{\xi}_i\in\mathcal F_i}{\arg\min}\left\{J_i(\boldsymbol{\xi}_i)+(\bar{\boldsymbol{\nu}}_i^\tau)^\top A_i\boldsymbol{\xi}_i+\frac{c}{2}\left\|A_i\boldsymbol{\xi}_i-A_i\boldsymbol{\xi}_i^\tau+\bar{\boldsymbol{\psi}}_i^\tau\right\|_2^2\right\}, \label{eq:tracking_primal}\\
\boldsymbol{\psi}_i^{\tau+1}&=\bar{\boldsymbol{\psi}}_i^\tau+A_i\boldsymbol{\xi}_i^{\tau+1}-A_i\boldsymbol{\xi}_i^\tau, \label{eq:tracking_tracker}\\
\boldsymbol{\nu}_i^{\tau+1}&=\bar{\boldsymbol{\nu}}_i^\tau+c\,\boldsymbol{\psi}_i^{\tau+1}. \label{eq:tracking_dual}
\end{align}
\end{subequations}

Here, $c>0$ is the constant penalty parameter. Equation~\eqref{eq:tracking_mix} performs neighbour-based mixing of the local dual and tracking variables, \eqref{eq:tracking_primal} solves the local convex MPC sub-problem, and \eqref{eq:tracking_tracker}--\eqref{eq:tracking_dual} update the local coupling-violation tracker and dual estimate. Each iteration therefore requires only local optimization and neighbour-to-neighbour communication \cite{falsone2020tracking}.

Under Assumptions~\ref{ass:convex_ccp} and~\ref{ass:tracking_admm}, Theorems~1 and~2 of \cite{falsone2020tracking} apply directly to \eqref{eq:tracking_admm}. The local tracking variables satisfy $\boldsymbol{\psi}_i^\tau\rightarrow\mathbf 0$, the local dual estimates converge to a common optimal dual solution, and any limit point of the primal sequence $\{(\boldsymbol{\xi}_1^\tau,\ldots,\boldsymbol{\xi}_N^\tau)\}_{\tau\ge0}$ is an optimal solution of \eqref{eq:general_ccp}. Hence, the Tracking-ADMM iteration asymptotically recovers an optimal solution of the centralized finite-horizon CCP using distributed local computation and neighbour communication.

\section{Systematic Feature Selection for Recursive Learning-Based MPC}
\label{sec:feature_selection}

\begin{figure*}[t!]
    \centering
    \includegraphics[width=0.9\textwidth]{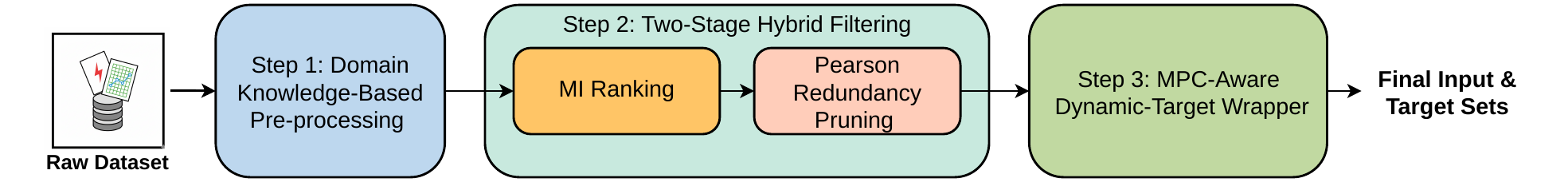}
    \caption{The proposed three-step feature-selection procedure for recursive learning-based MPC. Step 1 incorporates domain knowledge and classifies candidate features according to their future-value availability; Step 2 combines MI-based relevance ranking with Pearson redundancy pruning; and Step 3 applies the Dynamic-Target forward wrapper, under which selected future-unknown features are recursively predicted while selected horizon-known features remain input-only.}
    \label{fig:fs_flowchart}
\end{figure*}

The proposed methodology builds on the three-step filter-wrapper structure of Zhang and Wen \cite{zhang2019systematic}, while introducing MPC-oriented modifications to the pre-processing, filtering, and wrapper stages for recursive prediction. A key MPC-specific consideration is the availability of candidate-feature values over the prediction horizon. A selected auxiliary variable can remain input-only when its future trajectory is known, whereas a future-unknown auxiliary must also be predicted recursively if it is to remain available at subsequent prediction stages. The predictive benefit of each candidate is therefore assessed under recursive multi-step prediction rather than solely from one-step predictive improvement. The resulting procedure, illustrated in Fig.~\ref{fig:fs_flowchart}, combines domain-informed pre-processing, mutual-information (MI)-based relevance filtering and Pearson redundancy pruning, and an MPC-aware Dynamic-Target forward wrapper.

\subsection{Step 1: Domain Knowledge-Based Pre-processing}
\label{ssec:fs_step1}
The initial step refines the raw dataset by embedding essential domain knowledge. This involves four primary tasks:
\begin{itemize}[leftmargin=*]
    \item \textbf{Initial Pruning:} Raw features that are irrelevant to the intended control task or unavailable under realistic deployment conditions are removed from the candidate pool.
    \item \textbf{Feature Augmentation:} Physically or operationally meaningful derived features may be added to the candidate pool, including calendar-based variables that represent periodic operating patterns (e.g., hour of day, day of week).
    \item \textbf{Declaration of Mandatory Features:} The control inputs and primary prediction targets required by the MPC formulation are declared mandatory and are protected from removal in the subsequent filtering and wrapper stages.
    \item \textbf{Future Availability Classification:} Each non-mandatory candidate is classified according to whether its future trajectory is available over the MPC prediction horizon. A selected horizon-known candidate remains input-only, whereas a selected future-unknown candidate is included in both the model input set and the recursive target set so that its predicted value can be propagated to subsequent prediction stages.
\end{itemize}

\subsection{Step 2: Two-Stage Hybrid Filtering}
\label{ssec:fs_step2}
This step efficiently reduces the dimensionality of the candidate feature set. A key enhancement to the procedure in \cite{zhang2019systematic} is the use of a \textbf{two-stage hybrid filter}.
\begin{enumerate}[label=\arabic*), wide, labelindent=0pt]
    \item \textbf{MI Ranking:} First, the relevance of each non-mandatory candidate to the primary prediction targets is evaluated using MI. Whereas the implementation of Zhang and Wen \cite{zhang2019systematic} uses Pearson correlation to evaluate feature-output relevance, MI can capture nonlinear statistical dependence \cite{peng2005feature} and is therefore adopted for the relevance-ranking stage. For multiple primary targets, MI is computed separately between each candidate feature and each primary target. The resulting target-specific MI values are then combined into a single feature-level relevance score using prescribed weights. These weights allow greater emphasis to be placed on targets that are considered more important for the control task or more difficult to predict reliably. Only the highest-ranked fraction of non-mandatory candidates is retained, while mandatory controls and primary prediction targets remain protected.
    \item \textbf{Pearson Redundancy Pruning:} Second, absolute pairwise Pearson correlations are computed among the features retained after MI ranking. If the absolute correlation of a feature pair exceeds a prescribed threshold, the pair is considered redundant. Pearson correlation is used only to identify such redundant pairs, while the subsequent removal decision is based on feature protection and MI relevance. The removal rule is as follows: (i) if only one feature is protected as either a mandatory control or a primary prediction target, the unprotected feature is removed; (ii) if neither feature is protected, the feature with the lower MI score is removed; and (iii) if both features are protected, both are retained. Zhang and Wen \cite{zhang2019systematic} instead remove the member of a redundant pair that has the lower correlation with the model output. Here, the MI relevance score already obtained in the first filtering stage is used for this removal decision, while Pearson correlation is reserved for identifying redundancy.
\end{enumerate}

\subsection{Step 3: MPC-Aware Dynamic-Target Wrapper Selection}
\label{ssec:fs_step3}

The final stage applies a greedy forward wrapper using the IC-EoT architecture adopted for subsequent MPC deployment. At the initial round, the reference model contains only the mandatory control inputs and primary prediction targets. In each selection round, every remaining candidate is evaluated separately by temporarily adding it according to the future-availability rule defined in Section~\ref{ssec:fs_step1} and retraining the resulting trial model. Consequently, the recursive target set changes only when the evaluated candidate is future-unknown.

Each trial model is evaluated by a free-running $N_p$-step rollout on validation data. The rollout starts from the recorded historical input window. At each subsequent prediction stage, the recorded control input from the validation trajectory is supplied to the model, while all variables belonging to the recursive target set are replaced by the model predictions from the preceding stage. Recorded future values are retained only for selected horizon-known input-only features. Consequently, a future-unknown candidate is evaluated using its own recursively propagated predictions rather than unavailable future measurements.

The rollout score is evaluated only on the primary MPC targets. For each primary target and each prediction step, the model prediction and corresponding recorded target value are transformed using the standardization fitted to the training data, and their mean squared error (MSE) is then computed on this standardized scale. When multiple primary targets are considered, their standardized MSEs are first aggregated using prescribed weights, with larger weights allowed for targets that are considered more important for the control task or more difficult to predict reliably. The weighted error is then averaged over all prediction steps in the $N_p$-step rollout to produce a single score for the candidate model, with a lower score indicating better recursive predictive performance.

Prediction errors of selected auxiliary targets are not added separately to this score because feature selection is evaluated only on the primary prediction targets required by the MPC formulation. However, when a future-unknown auxiliary is selected, its predicted value is fed back into the model history at subsequent rollout steps. An inaccurate auxiliary prediction can therefore degrade later predictions of the primary targets, and this degradation is automatically reflected in their multi-step MSEs and hence in the rollout score.

Selection proceeds greedily. At the beginning of each round, the current input and recursive target sets define the reference model. Each remaining candidate is then added according to the future-availability rule in Section~\ref{ssec:fs_step1}, and a separate trial model is retrained and evaluated using the same recursive rollout procedure. The candidate yielding the lowest rollout score is identified, and its relative improvement with respect to the current reference model is computed from the reduction in rollout score normalized by the reference-model score. The candidate is accepted only when this relative improvement reaches a prescribed minimum threshold. If accepted, the feature is permanently added to the input set and, when future-unknown, also to the recursive target set; these updated sets define the reference model for the next round. If no candidate reaches the threshold, the forward-selection procedure terminates. The recursive target set can therefore evolve as features are selected, which motivates the term \textit{Dynamic-Target}.

\section{Case Study}
\label{sec:case_study}

This section presents a comprehensive empirical validation of the proposed distributed MPC framework. The case study addresses the DR problem for a heterogeneous building aggregation, where the optimization is formulated as a convex problem---a key contribution of this work---and solved using the Tracking-ADMM algorithm. To ensure a realistic and reproducible evaluation, a co-simulation environment is established, integrating the EnergyPlus building simulator \cite{crawley2001energyplus} with a Python-based controller via the Energym library \cite{scharnhorst2021energym}. The predictive models embedded within the MPC are IC-EoTs \cite{xu2026input}, whose input and output features are determined by the systematic selection methodology detailed in Section~\ref{sec:feature_selection}.

The performance of the proposed framework is rigorously evaluated through a series of comparative experiments designed to:
\begin{itemize}[leftmargin=*]
    \item Assess its ability to achieve near-optimal control performance by benchmarking against a theoretical centralized MPC controller.
    \item Demonstrate its computational scalability by comparing its solver time against the centralized approach as the size of the aggregation increases.
    \item Quantify the economic and operational benefits of coordinating a heterogeneous aggregation by comparing it against homogeneous groups and non-cooperative, individual optimization baselines.
\end{itemize}

\subsection{Simulation Framework and Building Testbeds}

\begin{figure*}[t!]
    \centering
    \includegraphics[width=0.9\textwidth]{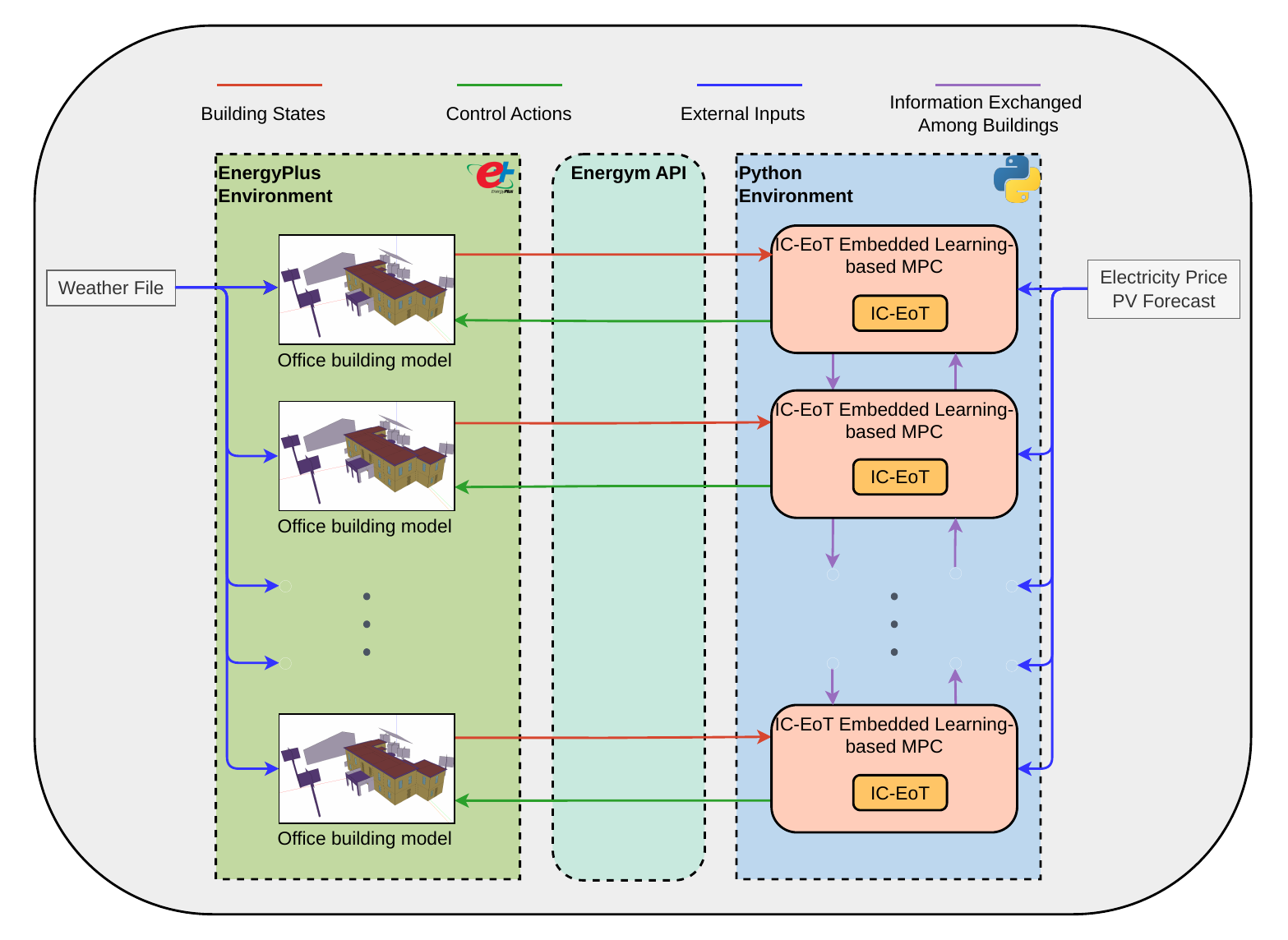}
    \caption{Closed-loop co-simulation architecture for the heterogeneous office-building aggregation. EnergyPlus building models communicate with their IC-EoT-embedded MPC agents through Energym. Weather information, electricity prices, and prosumer PV forecasts provide exogenous inputs, while neighbouring MPC agents exchange the tracking and dual information required for distributed coordination. The arrows distinguish building observations, control actions, exogenous inputs, and inter-agent communication.}
    \label{fig:sim_framework}
\end{figure*}

A unified EnergyPlus--Energym co-simulation framework is used for both predictive-model development and closed-loop aggregation-control experiments. Two Energym building environments are considered to evaluate the feature-selection methodology across distinct residential and commercial dynamics, while the subsequent distributed MPC experiments use the commercial office environment as the common thermal plant for all building agents. The building testbeds and their respective roles are introduced first, followed by the closed-loop aggregation architecture illustrated in Fig.~\ref{fig:sim_framework}.

Building dynamics are simulated using EnergyPlus \cite{crawley2001energyplus}, while Energym \cite{scharnhorst2021energym} provides the interface between the EnergyPlus models and the Python-based experimental pipeline. Through this interface, the controller receives building observations and weather information and returns thermostat control actions at each control update. The same interface is used both to generate simulated trajectories offline for predictive-model development and feature selection, and to execute the subsequent closed-loop MPC experiments.

The first testbed is the residential \texttt{ApartmentsThermal-v0} environment, which represents a four-storey building in Tarragona, Spain, with a total floor area of approximately $417.12~\mathrm{m}^2$ and an air volume of approximately $1042.83~\mathrm{m}^3$. The building contains eight thermal zones and four independently controllable thermostat setpoints \cite{scharnhorst2021energym}. The second testbed is the commercial \texttt{OfficesThermostat-v0} environment located in Greece, with a total floor area of approximately $643.73~\mathrm{m}^2$ and an air volume of approximately $2504.54~\mathrm{m}^3$. It contains 25 conditioned spaces, of which 14 thermal zones have independently controllable thermostat setpoints. Under the summer operating conditions considered here, the office building uses active mechanical cooling through an electric chiller \cite{scharnhorst2021energym}. The two environments therefore provide distinct building types, HVAC characteristics, and control dimensions for evaluating whether the feature-selection methodology generalizes beyond a single predictive-modelling task.

Figure~\ref{fig:sim_framework} illustrates the closed-loop architecture used for the aggregation-control experiments. Multiple instances of the office-building plant communicate with their corresponding IC-EoT-embedded MPC agents through Energym. Building observations are transmitted to the Python environment, while the resulting thermostat setpoints are returned to EnergyPlus as control actions. Electricity prices and prosumer PV forecasts are supplied as exogenous information to the MPC problems. For distributed coordination, neighbouring agents additionally exchange the local tracking and dual information required by Tracking-ADMM. These interactions are repeated at every MPC update to close the receding-horizon control loop.

The subsequent aggregation-control experiments use multiple instances of \texttt{OfficesThermostat-v0}. The agent set $\mathcal B$ introduced in Section~\ref{sec:methodology} is partitioned into $n_c$ consumer agents indexed by $\mathcal C$ and $n_p$ prosumer agents indexed by $\mathcal P$, such that
\begin{equation}
\mathcal B=\mathcal C\cup\mathcal P,\qquad \mathcal C\cap\mathcal P=\emptyset,\qquad N=n_c+n_p.
\label{eq:agent_partition}
\end{equation}
Both agent types share the same office-building thermal model and HVAC configuration. Consumer agents have no on-site generation and obtain electricity from the main grid or through internal trading within the aggregation. Prosumer agents are additionally associated with on-site PV generation and can allocate the available PV electricity among local consumption, export to the main grid, and internal trading. A common pre-trained PV forecasting model is used by all prosumer agents. Before each MPC update, the predictor is evaluated for each prosumer using its corresponding local input history to generate the PV trajectory over the complete prediction horizon. The resulting trajectory is subsequently treated as fixed exogenous information in the local prosumer MPC problem.

Accordingly, heterogeneity in the aggregation refers primarily to differences in distributed-energy-resource ownership and electricity-market participation rather than to different thermal-building archetypes. Individual agents are represented by distinct simulated operating trajectories and independently trained IC-EoT models based on their respective datasets. The resulting local predictive models, and hence the local MPC problems, are therefore not identical despite sharing the same office-building archetype.  Before these local optimization problems are formulated, the predictive variables and feature sets required by the controller must first be determined. The following subsection therefore applies the feature-selection methodology of Section~\ref{sec:feature_selection} to the two building testbeds and establishes the final IC-EoT representations used in the subsequent aggregation MPC.

\subsection{Predictive Model Development and Feature-Selection Validation}

Before formulating the consumer and prosumer MPC sub-problems, the predictive quantities required by the controller and the corresponding IC-EoT representations are established. The primary prediction targets are first determined from the subsequent control requirements, after which the feature-selection methodology in Section~\ref{sec:feature_selection} is instantiated and evaluated on both building testbeds introduced above. The resulting Office representation is then adopted for all subsequent aggregation-control experiments.

\subsubsection{Primary Prediction Targets and IC-EoT Configuration}

The predictive-model requirements are determined by the underlying building-control task. Thermostat setpoints are manipulated to influence facility electricity consumption while maintaining acceptable indoor temperatures. Accordingly, the primary prediction targets for the Office testbed are the 14 controlled zonal temperatures and total facility electricity consumption \texttt{Fa\_E\_All}, with the corresponding 14 thermostat setpoints treated as mandatory control inputs. For the Apartment testbed, the primary targets comprise eight zonal temperatures and \texttt{Fa\_E\_All}, with four thermostat setpoints serving as mandatory controls. These predictions subsequently provide the thermal and electrical quantities required by the MPC controller.

All predictive models use the IC-EoT formulation described in Section~\ref{sec:methodology}. The same model architecture and training protocol are retained throughout the feature-selection experiments so that differences in predictive performance can be attributed to the selected representation rather than to changes in model capacity or optimization settings.

All IC-EoT models use a fixed input-history window of 10 samples, one encoder layer, one attention head, a model dimension of 64, and a feed-forward dimension of 64, with no dropout. Softplus with a sharpness parameter of 20 is used throughout as the activation function, satisfying the required convex and non-decreasing structure of the IC-EoT formulation while providing smooth derivatives for gradient-based optimization. Training minimizes mean squared error using Adam with a learning rate of $10^{-4}$ and a batch size of 256, with a maximum of 2000 epochs and early stopping after 10 epochs without validation improvement. Input and output standardization parameters are fitted exclusively on the corresponding training data.

\subsubsection{Feature Selection Implementation and Results}

The methodology in Section~\ref{sec:feature_selection} is applied independently to the \texttt{OfficesThermostat-v0} and \texttt{ApartmentsThermal-v0} datasets. For both testbeds, separate simulated datasets are used for model training, feature-selection validation, and independent final testing. The training data are used for model fitting and MI evaluation, the validation data are used to compare candidate feature sets, and the independent test data are reserved exclusively for the final comparison of the selected representations.

Following domain-based pre-processing, the Office and Apartment candidate sets contain 61 and 59 variables, respectively. In the filtering stage, the MI ranking retains the highest-ranked 80\% of non-mandatory candidates, yielding 52 variables for Office and 51 for Apartment after the protected variables are retained. Pearson redundancy pruning with a threshold of 0.9 further reduces these sets to 46 and 30 variables, respectively. After excluding the mandatory controls and primary prediction targets, 17 auxiliary candidates remain for wrapper selection in each testbed.

The Dynamic-Target wrapper uses a prediction horizon of $N_p=8$, identical to that used in the subsequent MPC experiments, thereby evaluating candidate features over the same recursive horizon encountered during control. Facility-energy prediction and the mean contribution of the zonal-temperature targets are weighted by 0.6 and 0.4, respectively, in both the MI relevance score and wrapper evaluation metric. Forward selection terminates when the best remaining candidate reduces the rollout score by less than 0.5\%.

Three feature representations are compared for each testbed. The \textit{Core} representation contains only the mandatory controls and primary prediction targets. The \textit{Standard forward wrapper} adopts a conventional forward-selection pathway, as also used by Zhang and Wen \cite{zhang2019systematic}, but applies it to the common candidate pool produced by the MI and Pearson filtering stages in this study and evaluates each candidate using one-step weighted standardized MSE. The proposed \textit{Dynamic-Target} wrapper starts from the same filtered candidate pool but evaluates candidates using the free-running multi-step procedure in Section~\ref{ssec:fs_step3}. During Standard-wrapper selection, auxiliary candidates are treated as additional inputs only. For the final recursive comparison, any selected future-unknown auxiliary is additionally included in the model output set, ensuring that all representations can be rolled out without access to unavailable future measurements.

The auxiliary features selected by the two wrapper variants are summarized in Table~\ref{tab:fs_selected_features}.

\begin{table*}[t!]
\centering
\caption{Auxiliary features included in the compared feature representations.}
\label{tab:fs_selected_features}
\small
\begin{tabular}{llll}
\toprule
\textbf{Testbed} & \textbf{Representation} & \textbf{Future-unknown auxiliary features} & \textbf{Future-known input-only features}\\
\midrule
Office & Core & -- & --\\
& Standard forward wrapper & \begin{tabular}[t]{@{}l@{}}\texttt{Z25\_Fl\_Fan1\_sp\_out}, \texttt{Z17\_Fl\_Fan1\_sp\_out}\\\texttt{Z19\_Fl\_Fan1\_sp\_out}, \texttt{Z16\_Fl\_Fan\_sp\_out}\\\texttt{Z20\_Fl\_Fan1\_sp\_out}, \texttt{Z15\_Fl\_Fan\_sp\_out}\\\texttt{Z01\_Fl\_Fan\_sp\_out}\end{tabular} & --\\
& Proposed Dynamic-Target & \texttt{Ext\_T} & --\\
\midrule
Apartment & Core & -- & --\\
& Standard forward wrapper & \texttt{P2\_FlFrac\_HW}, \texttt{Fa\_Pw\_Prod}, \texttt{Z03\_RH} & --\\
& Proposed Dynamic-Target & \texttt{Fa\_E\_Light}, \texttt{Fa\_E\_HVAC} & \texttt{Hour}\\
\bottomrule
\end{tabular}
\end{table*}

For the Office testbed, the Dynamic-Target wrapper selects only the future-unknown feature \texttt{Ext\_T}. Its inclusion reduces the validation rollout score by 4.574\%, while the best remaining candidate in the following round provides only a 0.304\% improvement, below the prescribed 0.5\% threshold. For the Apartment testbed, \texttt{Fa\_E\_Light}, \texttt{Hour}, and \texttt{Fa\_E\_HVAC} are selected successively, reducing the validation rollout score from 0.4552 to 0.3368 before the stopping criterion is reached. Among these features, \texttt{Hour} is horizon-known and therefore remains input-only, whereas \texttt{Fa\_E\_Light} and \texttt{Fa\_E\_HVAC} are recursively predicted. After feature selection, the three final representations for each testbed are independently retrained using the common IC-EoT training protocol described above and are then evaluated on the independent test data using the same free-running recursive-rollout procedure.

\begin{figure*}[t]
\centering
\includegraphics[width=0.49\textwidth]{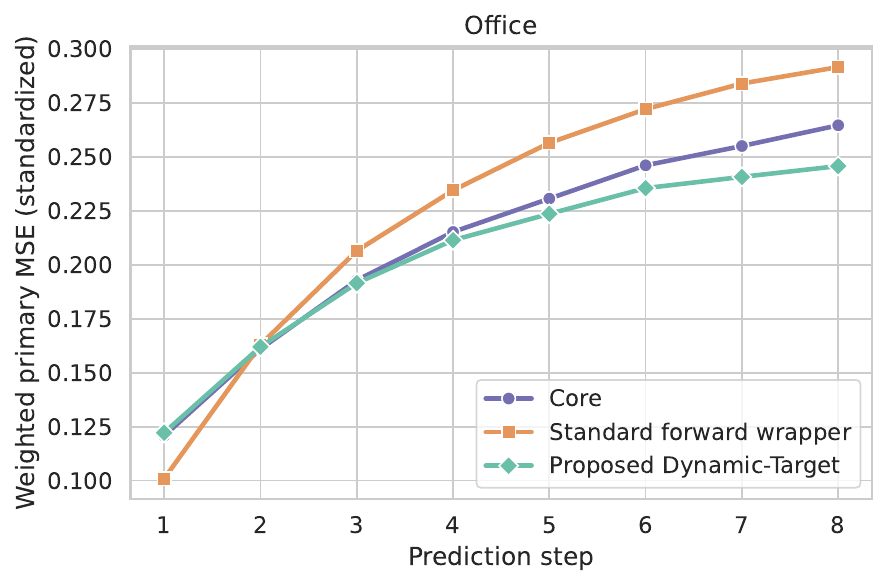}
\includegraphics[width=0.49\textwidth]{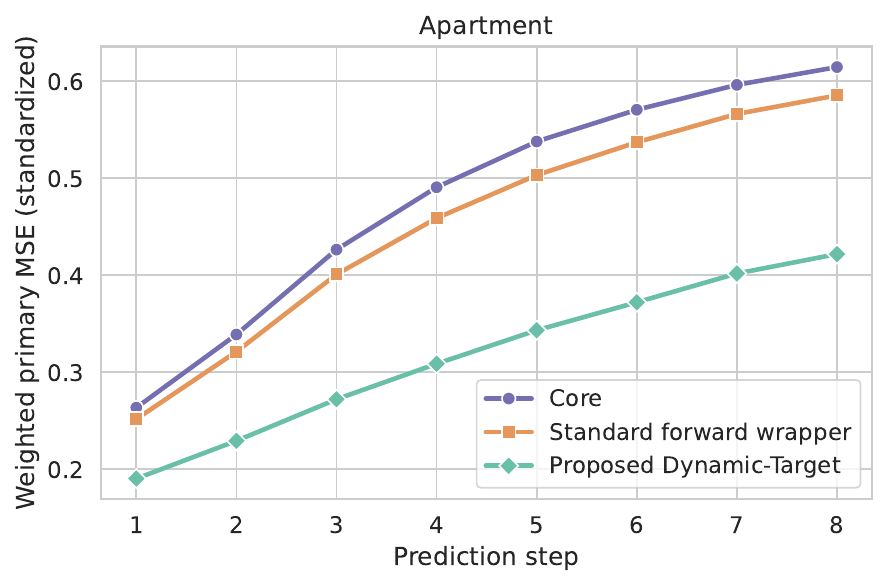}
\caption{Free-running recursive-rollout performance on independent test data for the Core, Standard forward-wrapper, and proposed Dynamic-Target representations over the eight-step prediction horizon: (a) Office testbed; (b) Apartment testbed. The weighted standardized MSE is evaluated on the common primary prediction targets using the same target weighting adopted by the proposed wrapper.}
\label{fig:fs_rollout_comparison}
\end{figure*}

\begin{table}[t!]
\centering
\caption{Quantitative summary of the weighted primary-target rollout MSE shown in Fig.~\ref{fig:fs_rollout_comparison}.}
\label{tab:fs_rollout_summary}
\small
\begin{tabular}{llccc}
\toprule
\textbf{Testbed} & \textbf{Representation} & \textbf{Step 1} & \textbf{Horizon mean} & \textbf{Step 8}\\
\midrule
Office & Core & 0.1208 & 0.2107 & 0.2646\\
& Standard & \textbf{0.1009} & 0.2261 & 0.2915\\
& Dynamic-Target & 0.1222 & \textbf{0.2040} & \textbf{0.2457}\\
\midrule
Apartment & Core & 0.2636 & 0.4799 & 0.6147\\
& Standard & 0.2518 & 0.4532 & 0.5855\\
& Dynamic-Target & \textbf{0.1904} & \textbf{0.3175} & \textbf{0.4219}\\
\bottomrule
\end{tabular}
\end{table}

Figure~\ref{fig:fs_rollout_comparison} and Table~\ref{tab:fs_rollout_summary} compare the recursive prediction performance of the three representations on independent test data. For the Office testbed, the proposed Dynamic-Target representation achieves the lowest mean weighted primary-target MSE over the eight-step horizon, reducing the corresponding value by 3.2\% relative to Core and by 9.7\% relative to the Standard forward wrapper. A particularly relevant difference appears across the prediction horizon: the Standard wrapper achieves the lowest error at the first step but the highest error at step 8, whereas the Dynamic-Target representation provides the lowest step-8 error.  This ranking reversal illustrates the limitation of relying solely on one-step predictive performance in this setting: features that improve one-step prediction may lose their predictive advantage over a recursive horizon as prediction errors accumulate.

The advantage is more pronounced for the Apartment testbed, where the Dynamic-Target representation reduces the horizon-mean weighted MSE by 33.8\% relative to Core and by 29.9\% relative to the Standard wrapper. It also maintains the lowest error throughout the reported prediction horizon, including at step 8. Together with the Office results, this indicates that explicitly accounting for future-value availability and recursive prediction during feature selection yields representations that are better aligned with the multi-step prediction requirements of MPC.

The Office representation identified by the proposed Dynamic-Target procedure is therefore adopted for all subsequent aggregation-control experiments. It uses the 14 thermostat setpoints as mandatory control inputs and recursively predicts the 14 controlled zonal temperatures, facility electricity consumption, and the selected future-unknown auxiliary variable \texttt{Ext\_T}. No additional horizon-known auxiliary feature is selected. The prosumer PV forecast remains external to the IC-EoT representation and is supplied directly to the prosumer MPC problem as fixed information over the prediction horizon.

\subsection{Heterogeneous Aggregation MPC and Distributed Implementation}
\label{sec:aggregation_mpc}

This subsection instantiates the general convex CCP of Section~\ref{sec:methodology} for the heterogeneous office-building aggregation under a Time-of-Use (TOU) demand response scheme and specifies its finite-iteration receding-horizon implementation. Consumers and prosumers share the same building thermal-control structure but differ in their available electricity sources and internal-market roles. Their local MPC problems minimize electricity expenditure under time-varying grid prices while enforcing HVAC actuation limits, thermal-comfort requirements, local electricity-supply constraints, prosumer PV-allocation constraints, and the aggregation-wide internal-market balance.

\subsubsection{Electricity Tariffs and Internal Trading}
The economic context for the optimization is defined by the grid-purchase tariff, the PV feed-in reference price, and the internal trading tariff. Consistent with the OfficesThermostat case study in~\cite{xu2026input}, the Public Power Corporation (PPC) $\Gamma 23$ Business tariff is adopted for grid electricity purchases, with the supply charges taken from the official PPC tariff schedule~\cite{PPC_G23_2026} and the corresponding summer dual-zone periods specified by the Hellenic Electricity Distribution Network Operator~\cite{HEDNO_DualZone_2025}. Accordingly, for each prediction stage $k\in\mathcal K$, the grid-purchase tariff is

\begin{equation}
\pi_{t+k}^{\mathrm{grid}}=
\begin{cases}
0.129~\text{\texteuro/kWh}, & 02{:}00\text{--}04{:}00\ \text{or}\ 11{:}00\text{--}15{:}00,\\
0.209~\text{\texteuro/kWh}, & \text{otherwise}.
\end{cases}
\label{eq:grid_tariff}
\end{equation}

The monthly fixed charge is excluded because it is independent of the control decisions and identical across the compared controllers.

For electricity exported by prosumers to the main grid, a fixed Greek photovoltaic feed-in tariff (FiT) reference price of
\begin{equation}
\pi_{\mathrm{FiT}}=0.063~\text{\texteuro/kWh}
\end{equation}
is adopted~\cite{HellenicMinistry_PV_FiT_2020}. This value corresponds to the reference price of $63~\text{\texteuro/MWh}$ for photovoltaic installations below $500~\mathrm{kW}$. The internal trading tariff (ITT) is defined at each prediction stage $k\in\mathcal K$ as
\begin{equation}
\pi_{t+k}^{\mathrm{ITT}}=\frac{\pi_{t+k}^{\mathrm{grid}}+\pi^{\mathrm{FiT}}}{2}.
\label{eq:itt_tariff}
\end{equation}

The resulting ordering $\pi^{\mathrm{FiT}}<\pi^{\mathrm{ITT}}_{t+k}<\pi^{\mathrm{grid}}_{t+k}$ provides an economic incentive for locally available PV surplus to be exchanged within the aggregation before equivalent electricity is purchased from or exported to the main grid.

\subsubsection{Consumer and Prosumer Local MPC Formulations}
\label{ssec:local_mpc}

The final Office representation identified in the preceding subsection is used by every aggregation agent. Each local IC-EoT recursively predicts the 14 controlled zone temperatures, facility electricity consumption \texttt{Fa\_E\_All}, and the selected future-unknown auxiliary variable \texttt{Ext\_T}. No horizon-known auxiliary feature is retained. Consequently, the deployed model has 16 recursively propagated outputs and, after applying the selective control extension in \eqref{eq:selective_extension}, a 44-dimensional input row comprising the 16 model variables, 14 thermostat controls, and their 14 negated copies. The recursive rollout follows \eqref{eq:history_update}--\eqref{eq:recursive_prediction}. Only the resulting temperature and facility-energy predictions enter the local MPC constraints, while \texttt{Ext\_T} is recursively propagated as an auxiliary model variable.

Thermal comfort follows the summer cooling formulation adopted in the IC-EoT study \cite{xu2026input} and the ICNN-based building MPC formulation of Bünning et al. \cite{bunning2021input}. Only an upper indoor-temperature constraint is imposed. From the convexity perspective, an upper bound on a convex predicted temperature defines a convex sublevel constraint, whereas an explicit lower-temperature bound would generally form a non-convex superlevel set \cite{bunning2021input}. In addition, unnecessary cooling is economically discouraged: for consumers it increases electricity purchases, while for prosumers it either increases grid purchases or reduces PV energy available for remunerated export or internal trading. The admissible thermostat-setpoint bounds further restrict excessive cooling. These mechanisms discourage overcooling but do not constitute a hard lower-temperature guarantee, consistent with the convex building-MPC formulations in \cite{xu2026input,bunning2021input}.

Following the formulations in \cite{xu2026input,bunning2021input}, the upper-temperature constraint is softened using non-negative slack variables to maintain feasibility in the presence of prediction errors or temporarily unavoidable overheating. Slack violations are penalized quadratically in the objective, while the finite upper bound $\boldsymbol{\epsilon}_{\max}$ prevents unrealistically large comfort relaxation and improves numerical conditioning. Because the predicted temperatures are convex in the control sequence and the slack variables enter affinely, the softened upper-temperature constraints remain convex.

For the numerical case study, the upper comfort limit is set to $24\,^{\circ}\mathrm{C}$ for all controlled zones, and every thermostat setpoint is constrained to $[16,26]\,^{\circ}\mathrm{C}$. All electricity-flow variables below denote energy over one 15-min control interval and are expressed in kWh. The quantity $\hat{\mathbf T}_{k+1|t}^i(\mathbf U_t^i)$ collects the 14 predicted controlled-zone temperatures, while $\hat E_{\mathrm{fac},k+1|t}^i(\mathbf U_t^i)$ denotes the predicted facility electricity consumption corresponding to \texttt{Fa\_E\_All}. The PV conversion efficiency is fixed at $\eta_{\mathrm{conv}}=0.95$. In addition to the comfort-slack penalty weighted by $\lambda_\epsilon$, a small positive coefficient $\lambda_{\mathrm{tr}}$ regularizes internal energy exchange. This term discourages economically unnecessary internal trading and reduces degeneracy among otherwise equivalent trading allocations. It may also be viewed as a lumped representation of small transaction or transfer overheads associated with internal exchange; explicit electrical network losses are not modelled.

\vspace{1ex}
\noindent\textbf{Consumer Agent Sub-problem}
\vspace{1ex}

For a consumer agent $i\in\mathcal C$, the local decisions comprise the 14-dimensional thermostat-setpoint sequence $\mathbf U_t^i$, the grid-purchase trajectory $\mathbf E_{\mathrm{grid},t}^i$, the internal-market purchase trajectory $\mathbf E_{\mathrm{agg},t}^i$, and the temperature-slack trajectory $\boldsymbol{\epsilon}_t^i$. At prediction stage $k$, $E_{\mathrm{grid},k|t}^i$ denotes electricity purchased from the main grid and $E_{\mathrm{agg},k|t}^i$ denotes electricity obtained from prosumers through the internal aggregation market. The objective minimizes finite-horizon electricity expenditure while penalizing unnecessary internal exchange and upper-temperature violations. The consumer interacts with other agents only through its internal-market purchase trajectory. The following constraints apply at every prediction stage $k\in\mathcal K$.

% ---------------------- Consumer Sub-Problem ----------------------
\begin{equation}
\boldsymbol{\xi}_t^i=
\operatorname{col}\!\left(
\mathbf U_t^i,
\mathbf E_{\mathrm{grid},t}^i,
\mathbf E_{\mathrm{agg},t}^i,
\boldsymbol{\epsilon}_t^i
\right),\qquad i\in\mathcal C.
\end{equation}

\begin{figure*}[!t]
\hrulefill
\begin{subequations}\label{eq:consumer_subproblem}
\begin{align}
\min_{\boldsymbol{\xi}_t^i}\quad J_t^i(\boldsymbol{\xi}_t^i):=&\sum_{k\in\mathcal K}\left[\pi_{t+k}^{\mathrm{grid}}E_{\mathrm{grid},k|t}^i+\pi_{t+k}^{\mathrm{ITT}}E_{\mathrm{agg},k|t}^i+\lambda_{\mathrm{tr}}E_{\mathrm{agg},k|t}^i+\lambda_\epsilon\|\boldsymbol{\epsilon}_{k+1|t}^i\|_2^2\right], \label{eq:consumer_obj_sub}\\
\text{s.t.}\quad &\hat{\mathbf T}_{k+1|t}^i(\mathbf U_t^i)-\boldsymbol{\epsilon}_{k+1|t}^i\preceq\mathbf T_{\max}, \label{eq:consumer_temp_sub}\\
&\hat E_{\mathrm{fac},k+1|t}^i(\mathbf U_t^i)\le E_{\mathrm{grid},k|t}^i+E_{\mathrm{agg},k|t}^i, \label{eq:consumer_balance_sub}\\
&\mathbf u_{\min}\preceq\mathbf u_{t+k|t}^i\preceq\mathbf u_{\max}, \label{eq:consumer_act_sub}\\
&\mathbf 0\preceq\boldsymbol{\epsilon}_{k+1|t}^i\preceq\boldsymbol{\epsilon}_{\max}, \label{eq:consumer_slack_sub}\\
&E_{\mathrm{grid},k|t}^i\ge0,\qquad E_{\mathrm{agg},k|t}^i\ge0. \label{eq:consumer_energy_bounds}
\end{align}
\end{subequations}
\hrulefill
\end{figure*}

\vspace{1ex}
\noindent\textbf{Prosumer Agent Sub-problem}
\vspace{1ex}

A prosumer agent $j\in\mathcal P$ shares the same thermal prediction and HVAC-control structure as a consumer but additionally has on-site PV generation. All prosumer agents use the same pre-trained PV forecasting model. Before each MPC solve, the model uses the available PV-power history to generate the forecast over the complete MPC prediction horizon. Forecast generation is performed before the local MPC optimization, so the resulting PV trajectory enters each local prosumer problem as fixed exogenous information, as illustrated in Fig.~\ref{fig:sim_framework}. Because all prosumer instances use the same weather file and PV configuration, they experience the same PV-generation profile and consequently receive the same horizon forecast. This removes variation in solar availability across prosumers, allowing the case study to focus on market-role heterogeneity and distributed coordination.

\begin{equation}
\bar{\mathbf E}_{\mathrm{PV},t}^j=\operatorname{col}\!\left(\bar E_{\mathrm{PV},0|t}^j,\ldots,\bar E_{\mathrm{PV},N_p-1|t}^j\right),
\label{eq:pv_forecast}
\end{equation}
which remains fixed throughout the corresponding optimization. The PV forecasting model is therefore external to the IC-EoT optimization mapping and is not required to satisfy the input-convexity conditions of Section~\ref{sec:methodology}.

At each prediction stage, $E_{\mathrm{PV}\to\mathrm L,k|t}^j$, $E_{\mathrm{PV}\to\mathrm G,k|t}^j$, and $E_{\mathrm{PV}\to\mathrm{Agg},k|t}^j$ denote the pre-conversion PV energy allocated to local building consumption, export to the main grid, and export to the internal aggregation market, respectively, while $E_{\mathrm G\to\mathrm L,k|t}^j$ denotes electricity imported from the main grid to supply the remaining local demand. The corresponding usable alternating-current (AC)-side PV quantities are obtained through the conversion efficiency $\eta_{\mathrm{conv}}$ defined above. The following constraints apply at every prediction stage $k\in\mathcal K$.

% ---------------------- Prosumer Sub-Problem ----------------------
\begin{equation}
\boldsymbol{\xi}_t^j=
\operatorname{col}\!\left(
\mathbf U_t^j,
\mathbf E_{\mathrm{PV}\to\mathrm L,t}^j,
\mathbf E_{\mathrm{PV}\to\mathrm G,t}^j,
\mathbf E_{\mathrm{PV}\to\mathrm{Agg},t}^j,
\mathbf E_{\mathrm G\to\mathrm L,t}^j,
\boldsymbol{\epsilon}_t^j
\right).
\label{eq:prosumer_decision}
\end{equation}

\begin{figure*}[!t]
\hrulefill
\begin{subequations}\label{eq:prosumer_subproblem}
\begin{align}
\min_{\boldsymbol{\xi}_t^j}\quad J_t^j(\boldsymbol{\xi}_t^j):=&\sum_{k\in\mathcal K}\Big[\pi_{t+k}^{\mathrm{grid}}E_{\mathrm G\to\mathrm L,k|t}^j-\pi^{\mathrm{FiT}}\eta_{\mathrm{conv}}E_{\mathrm{PV}\to\mathrm G,k|t}^j-\pi_{t+k}^{\mathrm{ITT}}\eta_{\mathrm{conv}}E_{\mathrm{PV}\to\mathrm{Agg},k|t}^j+\lambda_{\mathrm{tr}}\eta_{\mathrm{conv}}E_{\mathrm{PV}\to\mathrm{Agg},k|t}^j+\lambda_\epsilon\|\boldsymbol{\epsilon}_{k+1|t}^j\|_2^2\Big], \label{eq:prosumer_obj_sub}\\
\text{s.t.}\quad &\hat{\mathbf T}_{k+1|t}^j(\mathbf U_t^j)-\boldsymbol{\epsilon}_{k+1|t}^j\preceq\mathbf T_{\max}, \label{eq:prosumer_temp_sub}\\
&\bar E_{\mathrm{PV},k|t}^j=E_{\mathrm{PV}\to\mathrm L,k|t}^j+E_{\mathrm{PV}\to\mathrm G,k|t}^j+E_{\mathrm{PV}\to\mathrm{Agg},k|t}^j, \label{eq:prosumer_pv_alloc_sub}\\
&\hat E_{\mathrm{fac},k+1|t}^j(\mathbf U_t^j)\le E_{\mathrm G\to\mathrm L,k|t}^j+\eta_{\mathrm{conv}}E_{\mathrm{PV}\to\mathrm L,k|t}^j, \label{eq:prosumer_load_bal_sub}\\
&\mathbf u_{\min}\preceq\mathbf u_{t+k|t}^j\preceq\mathbf u_{\max}, \label{eq:prosumer_act_sub}\\
&\mathbf 0\preceq\boldsymbol{\epsilon}_{k+1|t}^j\preceq\boldsymbol{\epsilon}_{\max}, \label{eq:prosumer_slack_sub}\\
&E_{q,k|t}^j\ge0,\qquad q\in\{\mathrm{PV}\!\to\!\mathrm L,\mathrm{PV}\!\to\!\mathrm G,\mathrm{PV}\!\to\!\mathrm{Agg},\mathrm G\!\to\!\mathrm L\}. \label{eq:prosumer_flow_bounds}
\end{align}
\end{subequations}
\hrulefill
\end{figure*}

For accounting, grid and internal-market PV exports are evaluated on the AC side as $\eta_{\mathrm{conv}}E_{\mathrm{PV}\to\mathrm G,k|t}^j$ and $\eta_{\mathrm{conv}}E_{\mathrm{PV}\to\mathrm{Agg},k|t}^j$, respectively.

The load-supply relations in \eqref{eq:consumer_balance_sub} and \eqref{eq:prosumer_load_bal_sub} are deliberately formulated as inequalities. Imposing equality with the recursively substituted IC-EoT facility energy prediction would introduce a non-affine equality involving a convex nonlinear function and would therefore destroy the convex formulation established in Section~\ref{sec:methodology}. For consumers, purchasing unnecessary electricity increases the local cost because both grid and internal-trading prices are positive. For prosumers, unnecessary grid purchases increase cost, while allocating additional PV to local consumption reduces the amount available for remunerated grid export or internal trading. Consequently, provided that the available export channels are not simultaneously saturated, an optimum can be selected for which the intended load-supply balance is active without explicitly imposing a nonlinear equality constraint.

In contrast, the PV-allocation equality \eqref{eq:prosumer_pv_alloc_sub} is affine because the PV forecast is fixed before optimization. Explicit PV curtailment is not modelled; the predicted PV energy is allocated among local consumption, grid export, and internal trading.

\vspace{1ex}
\noindent\textbf{Aggregation-Wide Coupling Constraint}
\vspace{1ex}

For every prediction stage $k\in\mathcal K$, the only inter-agent constraint is the clearing condition of the internal electricity market:

\begin{equation}
\sum_{j\in\mathcal P}\eta_{\mathrm{conv}}E_{\mathrm{PV}\to\mathrm{Agg},k|t}^j-\sum_{i\in\mathcal C}E_{\mathrm{agg},k|t}^i=0.
\label{eq:internal_market_balance}
\end{equation}

Stacking \eqref{eq:internal_market_balance} over the prediction horizon yields
\begin{equation}
\sum_{i\in\mathcal B}A_i\boldsymbol{\xi}_t^i=\mathbf 0,
\end{equation}

which is the case-study realization of \eqref{eq:ccp_coupling} with $\mathbf b_t=\mathbf 0$. For a consumer, $A_i\boldsymbol{\xi}_t^i$ extracts the negative internal-purchase sequence, whereas for a prosumer it extracts the positive AC-side internal-export sequence $\eta_{\mathrm{conv}}\mathbf E_{\mathrm{PV}\to\mathrm{Agg},t}^i$.

\subsubsection{Practical Finite-Iteration Distributed MPC Implementation}

The local MPC problems in Section~\ref{ssec:local_mpc}, together with the affine internal-market clearing constraint, instantiate the convex CCP formulated in Section~\ref{sec:methodology}. Section~\ref{sec:methodology} establishes the asymptotic Tracking-ADMM solution of this problem class. In online MPC, however, only a finite computation window is available between successive control updates, so each distributed optimization must terminate after a finite number of Tracking-ADMM iterations. The implementation below therefore retains the primal, tracking, and dual updates in \eqref{eq:tracking_admm}, while introducing communication and termination mechanisms tailored to finite-time receding-horizon operation.

Let $\mathcal G$ denote the fixed undirected communication graph. A prescribed upper bound is imposed on the number of neighbours of each agent. When all other agents fall within this bound, $\mathcal G$ is complete; otherwise, a connected regular graph is used at the prescribed degree. Consequently, once the aggregation exceeds the small complete-graph case, the number of direct communication neighbours of each building remains fixed as additional buildings are introduced.

\begin{equation}
m_{ij}=
\begin{cases}
\dfrac{1}{2}, & j=i,\\[1mm]
\dfrac{1}{2(|\mathcal N_i|-1)}, & j\in\mathcal N_i\setminus\{i\},\\[1mm]
0, & \text{otherwise}.
\end{cases}
\label{eq:lazy_sparse_mixing}
\end{equation}

For the regular undirected graphs considered here, \eqref{eq:lazy_sparse_mixing} yields a symmetric and doubly stochastic mixing matrix. The resulting matrix is also positive semidefinite and therefore satisfies the communication conditions in Assumption~\ref{ass:tracking_admm}.

Let $R_{\mathrm{mix}}\ge1$ denote the number of neighbour-mixing rounds performed within each Tracking-ADMM iteration.

\begin{equation}
\bar{\boldsymbol{\nu}}_i^\tau=\sum_{j\in\mathcal B}[\mathbf M^{R_{\mathrm{mix}}}]_{ij}\boldsymbol{\nu}_j^\tau,\qquad
\bar{\boldsymbol{\psi}}_i^\tau=\sum_{j\in\mathcal B}[\mathbf M^{R_{\mathrm{mix}}}]_{ij}\boldsymbol{\psi}_j^\tau.
\label{eq:repeated_mixing}
\end{equation}

Equation~\eqref{eq:repeated_mixing} is implemented through $R_{\mathrm{mix}}$ consecutive one-hop neighbour exchanges; $\mathbf M^{R_{\mathrm{mix}}}$ is only the equivalent compact representation of these successive local exchanges. Since $\mathbf M$ is symmetric, doubly stochastic, and positive semidefinite, $\mathbf M^{R_{\mathrm{mix}}}$ retains these properties and remains compatible with Assumption~\ref{ass:tracking_admm}.

The repeated exchanges accelerate the propagation of tracking and dual information over the sparse graph before the next local MPC problem is solved. It therefore trades additional lightweight neighbour communication for the possibility of reducing the number of comparatively expensive local optimization and Tracking-ADMM iterations required for coordination.

Successive MPC problems use the shifted primal and dual trajectories from the preceding receding-horizon solution as warm starts, avoiding repeated initialization from an uncoordinated trajectory. The tracking variable is not shifted from the preceding MPC problem. Instead, after the current primal initialization has been constructed, it is reinitialized according to the Tracking-ADMM requirement $\boldsymbol{\psi}_i^0=A_i\boldsymbol{\xi}_i^0-\mathbf b_i$. Warm starting therefore changes only the numerical initialization of successive solves and does not modify the finite-horizon optimization problem or the Tracking-ADMM updates.

Because each MPC update is subject to a finite online computation budget, the distributed iterations must be terminated using finite-iteration criteria rather than waiting for asymptotic convergence. The stopping rule is therefore constructed from quantities available to each agent through its own variables and neighbour exchanges.

Let $\varepsilon_{\mathrm{feas}}>0$ denote the local coupling-feasibility tolerance and $\varepsilon_{\mathrm{cons}}>0$ the neighbour-consensus tolerance. Agent $i$ declares itself ready at iteration $\tau$ when

\begin{equation}
\mathrm{ready}_i^\tau=1
\Longleftrightarrow
\begin{cases}
\|\boldsymbol{\psi}_i^\tau\|_\infty\le\varepsilon_{\mathrm{feas}},\\[1mm]
\displaystyle\max_{j\in\mathcal N_i}\|\boldsymbol{\psi}_i^\tau-\boldsymbol{\psi}_j^\tau\|_\infty\le\varepsilon_{\mathrm{cons}},\\[2mm]
\displaystyle\max_{j\in\mathcal N_i}\left\|\dfrac{\boldsymbol{\nu}_i^\tau-\boldsymbol{\nu}_j^\tau}{c}\right\|_\infty\le\varepsilon_{\mathrm{cons}}.
\end{cases}
\label{eq:local_readiness}
\end{equation}

Let $\mathrm{ready}_{\mathcal B}^{\tau}$ denote the network-wide readiness flag, which is true only when all local flags $\mathrm{ready}_i^\tau$, $i\in\mathcal B$, are true. This flag is obtained without a central coordinator. Each agent repeatedly exchanges its current Boolean result with its direct neighbours and updates it using logical AND. Because $\mathcal G$ is connected, after $\operatorname{diam}(\mathcal G)$ such communication rounds, readiness information from every agent has reached every other agent, and all agents therefore obtain the same value of $\mathrm{ready}_{\mathcal B}^{\tau}$.

Let $\tau_{\max}$ denote the maximum number of Tracking-ADMM iterations permitted within one MPC update. The distributed solve terminates when network-wide readiness is established or when $\tau=\tau_{\max}$ is reached.

% ---------------------- Cross-page algorithm block ----------------------
\refstepcounter{algorithm}\label{alg:tracking_admm_impl}
\begin{tcolorbox}[breakable,enhanced,
  title={Algorithm~\thealgorithm: Finite-Iteration Tracking-ADMM Implementation per Agent $i$}]
{\small
\begin{algorithmic}[1]

\Require Penalty parameter $c>0$; communication graph $\mathcal G$ and mixing matrix $\mathbf M$; mixing rounds $R_{\mathrm{mix}}$; tolerances $\varepsilon_{\mathrm{feas}}$ and $\varepsilon_{\mathrm{cons}}$; iteration budget $\tau_{\max}$.

\Statex
\State \textbf{Initialize:} Initialize $\boldsymbol{\xi}_i^0$ and $\boldsymbol{\nu}_i^0$, using the shifted preceding MPC solution as a warm start when available.
\State Set $\boldsymbol{\psi}_i^0\gets A_i\boldsymbol{\xi}_i^0-\mathbf b_i$.

\For{$\tau=0,\ldots,\tau_{\max}-1$}

    \Statex \hspace{\algorithmicindent}\textbf{Step 1: Repeated Neighbour Mixing}
    \State Perform $R_{\mathrm{mix}}$ consecutive neighbour-mixing rounds according to \eqref{eq:repeated_mixing}.

    \Statex \hspace{\algorithmicindent}\textbf{Step 2: Local Convex Optimization}
    \State Solve the local convex MPC sub-problem according to \eqref{eq:tracking_primal}.

    \Statex \hspace{\algorithmicindent}\textbf{Step 3: Tracking and Dual Update}
    \State Update $\boldsymbol{\psi}_i^{\tau+1}$ and $\boldsymbol{\nu}_i^{\tau+1}$ according to \eqref{eq:tracking_tracker}--\eqref{eq:tracking_dual}.

    \Statex \hspace{\algorithmicindent}\textbf{Step 4: Fully Distributed Termination}
    \State Evaluate $\mathrm{ready}_i^{\tau+1}$ using \eqref{eq:local_readiness}.
    \State Perform $\operatorname{diam}(\mathcal G)$ rounds of neighbour-wise logical-AND communication to obtain $\mathrm{ready}_{\mathcal B}^{\tau+1}$.
    \If{$\mathrm{ready}_{\mathcal B}^{\tau+1}=1$}
        \State \textbf{break}
    \EndIf

\EndFor

\State Apply the first thermostat-setpoint vector of the resulting horizon solution.

\end{algorithmic}
}
\end{tcolorbox}
% -------------------- End of cross-page algorithm block ----------------------

Algorithm~\ref{alg:tracking_admm_impl} therefore preserves the local Tracking-ADMM optimization and update structure of Section~\ref{sec:methodology}, while combining bounded-neighbour communication, repeated low-cost information mixing, and finite fully distributed termination for receding-horizon MPC implementation. The specific numerical settings of these implementation elements are reported with the corresponding experiments in the following subsections.

\subsection{Computational Scalability and Convergence}
\label{ssec:scalability_convergence}

The proposed distributed implementation is evaluated from two complementary perspectives. First, its computational scalability is examined as the number of coordinated buildings increases, together with the resulting closed-loop economic and comfort performance. Second, for aggregation sizes for which a centralized benchmark remains computationally tractable, the finite-iteration Tracking-ADMM solution is compared directly with the corresponding centralized solution on the same MPC problem. This separation distinguishes closed-loop control performance from convergence of the underlying distributed optimization.

Unless otherwise stated, the remaining MPC experiments use the same numerical settings: $c=0.05$, $R_{\mathrm{mix}}=10$, $\varepsilon_{\mathrm{feas}}=\varepsilon_{\mathrm{cons}}=5\times10^{-3}$~kWh, $\tau_{\max}=40$, and an IPOPT convergence tolerance of $10^{-4}$. All timing experiments are performed on the same workstation equipped with an AMD Ryzen Threadripper PRO 7965WX processor (24 physical cores and 48 logical CPUs) and 125~GiB of memory. Each experimental process was assigned to a single logical CPU, and the numerical libraries were restricted to one computational thread.

\subsubsection{Computational Scalability and Closed-Loop Performance}
\label{sssec:scalability}

Scalability is evaluated using symmetric heterogeneous aggregations with $n_c=n_p$. The total aggregation size $N=n_c+n_p$ is increased from 2 to 14 buildings, corresponding to $n_c=n_p=1,\ldots,7$. The distributed controller uses the bounded-neighbour topology, repeated mixing, and fully distributed stopping mechanism described in Section~\ref{sec:aggregation_mpc}.

A PV-generation multiplier of 2.5 is applied to all prosumers in the scalability experiments. The native PV profile provides only limited instantaneous surplus after local consumption, leaving little energy available for internal exchange. The adopted scaling is therefore used to ensure that the coordination mechanism is exercised under meaningful prosumer surplus conditions, while leaving the building dynamics, HVAC configuration, and weather conditions unchanged. Across the evaluated cases, the resulting surplus is typically on the order of 10--17\% of the prosumers' own electricity demand, indicating that PV surplus remains a secondary rather than dominant source of energy. A higher multiplier of 4.0 is considered only in the subsequent coordination-value sensitivity study, where the corresponding surplus increases to approximately 30--40\% of prosumer demand, allowing the effect of greater redistributable PV availability to be examined explicitly.

For CMPC, the reported computation time is the wall-clock time required to solve the joint centralized optimization problem at each MPC update. For DMPC, the local MPC problems are distributable across agents and are therefore treated as concurrent computations. Its online computation time is evaluated by accumulating the local optimization time of each agent over the Tracking-ADMM iterations within one MPC update and taking the maximum across agents. This provides the parallel-equivalent computation time of the distributed implementation.

\begin{figure*}[t!]
\centering
\includegraphics[width=0.48\textwidth]{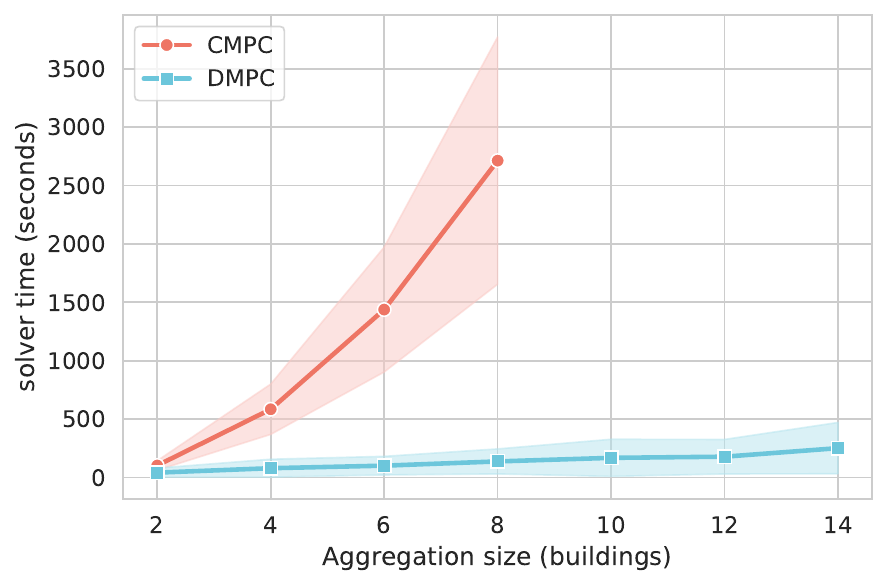}
\hfill
\includegraphics[width=0.48\textwidth]{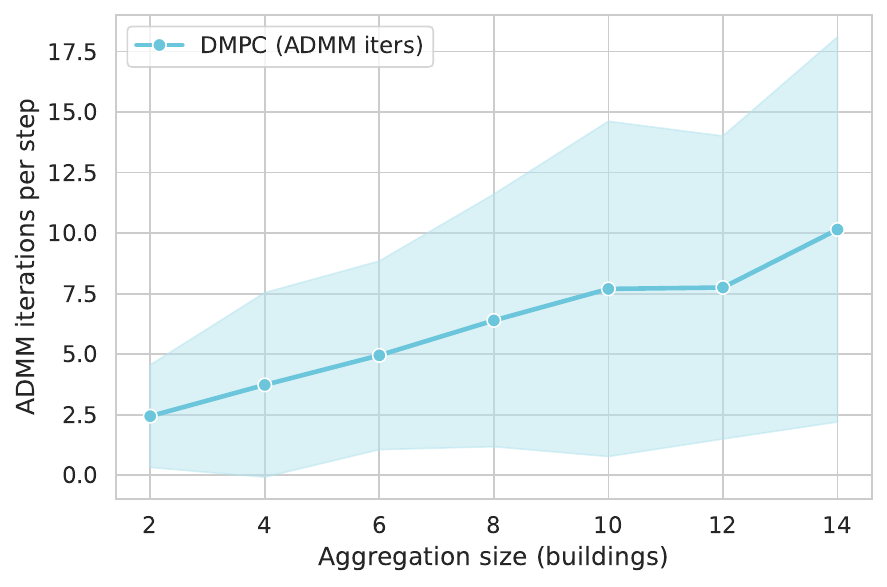}
\caption{Computational scalability with increasing aggregation size. Left: mean CMPC solution time and parallel-equivalent DMPC solution time per MPC update. Right: mean number of Tracking-ADMM iterations per MPC update. Shaded regions indicate one standard deviation.}
\label{fig:scalability}
\end{figure*}

Figure~\ref{fig:scalability} shows a pronounced separation in computational scaling between the centralized and distributed formulations. The mean CMPC solution time grows rapidly as the joint optimization problem expands with $N$. It remains below the 15-min control interval for $N=2$ and 4, exceeds this interval at $N=6$, and increases further at $N=8$. The centralized scalability sweep is therefore terminated at $N=8$, since the transition from online-feasible to clearly impractical computation has already been captured.

The DMPC parallel-equivalent computation time grows substantially more gradually and remains well below the 15-min control interval throughout the tested range up to $N=14$. Increasing $N$ does not enlarge the dimension of any individual building's local IC-EoT-embedded MPC problem; the size-dependent increase in computation instead arises primarily from additional coordination iterations. This interpretation is supported by comparing the two panels of Fig.~\ref{fig:scalability}: the increase in DMPC solution time in the left panel is accompanied by a rising mean Tracking-ADMM iteration count in the right panel, while the dimension of each local optimization problem remains unchanged. The distributed decomposition therefore converts the rapid growth of one centralized optimization problem into an increase primarily in the number of fixed-size local optimization rounds. Extending the distributed sweep beyond $N=14$ was not considered necessary, since the tested range already establishes a substantial computational separation from CMPC while keeping the full closed-loop simulation campaign tractable.

Table~\ref{tab:scalability_performance} shows that the finite-iteration distributed controller retains closely comparable closed-loop performance over the aggregation sizes for which a centralized reference is available. The electricity-bill difference remains within approximately $\pm1.2\%$ for $N=2$--8. The average upper-temperature violation is slightly higher under DMPC in these cases, but the absolute differences remain small, with the largest difference being approximately $0.051~^\circ\mathrm{C}\!\cdot\mathrm{h}$ per zone. DMPC additionally completes the full closed-loop experiments for $N=10$, 12, and 14, for which centralized closed-loop simulations are not pursued because of their prohibitive solution times.

\begin{table*}[t!]
\centering
\caption{Closed-loop performance of CMPC and DMPC across the scalability experiments.}
\label{tab:scalability_performance}
\small
\begin{tabular}{c ccc ccc}
\toprule
& \multicolumn{3}{c}{\textbf{Total electricity bill (\texteuro)}} & \multicolumn{3}{c}{\textbf{Average temperature violation ($^\circ$C$\cdot$h/zone)}}\\
\cmidrule(lr){2-4}\cmidrule(lr){5-7}
\textbf{$N$ ($n_c+n_p$)} & \textbf{CMPC} & \textbf{DMPC} & \textbf{$\Delta$ (\%)} & \textbf{CMPC} & \textbf{DMPC} & \textbf{$\Delta$ (\%)}\\
\midrule
2 (1+1) & 21.67 & 21.65 & -0.11 & 0.889 & 0.923 & +3.83\\
4 (2+2) & 42.75 & 42.48 & -0.62 & 0.771 & 0.789 & +2.29\\
6 (3+3) & 64.15 & 63.38 & -1.20 & 0.608 & 0.659 & +8.40\\
8 (4+4) & 83.73 & 84.45 & +0.86 & 0.594 & 0.623 & +4.99\\
10 (5+5) & -- & 105.45 & -- & -- & 0.682 & --\\
12 (6+6) & -- & 128.80 & -- & -- & 0.591 & --\\
14 (7+7) & -- & 148.50 & -- & -- & 0.600 & --\\
\bottomrule
\end{tabular}
\end{table*}

The small closed-loop differences between CMPC and DMPC should not be interpreted directly as optimization error at an individual MPC update. Tracking-ADMM is terminated according to finite stopping tolerances and an iteration budget, while the underlying local convex nonlinear programs are themselves solved to finite numerical tolerances. Small differences in the first applied control action can therefore produce slightly different subsequent plant states; from the next control update onward, the controllers are no longer solving identical MPC instances, and these differences can accumulate along the closed-loop trajectories. Accordingly, Table~\ref{tab:scalability_performance} establishes comparable closed-loop control performance but does not by itself demonstrate convergence to the centralized optimum. Section~\ref{sssec:convergence} therefore performs an optimization-level comparison at the same MPC update and with identical problem data, examining objective deviation, coupling feasibility, tracking disagreement, and dual disagreement across the Tracking-ADMM iterations.

\subsubsection{Convergence to the Centralized Solution}
\label{sssec:convergence}

To isolate optimization convergence from the closed-loop trajectory divergence discussed above, the comparison is performed at the first MPC update following the common history initialization for the $N=2$, 4, 6, and 8 symmetric aggregations. At this update, CMPC and DMPC are constructed from identical building histories, electricity prices, PV forecasts, predictive models, local constraints, and local objective functions, and therefore target the same underlying finite-horizon CCP. Their numerical solution procedures are nevertheless different: CMPC solves the joint CCP directly, whereas Tracking-ADMM iteratively solves the augmented local sub-problems in \eqref{eq:tracking_primal} and coordinates them through the tracking and dual updates. The convergence study considers all aggregation sizes for which the corresponding centralized benchmark simulations were completed. Although the mean CMPC solution time already exceeds the 15-min control interval at $N=6$, the $N=6$ and $N=8$ cases are retained as offline optimization benchmarks; larger centralized cases were not pursued because of their substantially increased simulation cost.

Four quantities are examined over the Tracking-ADMM iterations. Following the optimality assessment used in \cite{falsone2020tracking}, the first is the absolute relative deviation between the original CCP objective evaluated at the distributed primal iterate and the corresponding centralized optimum. The dual and quadratic penalty terms appearing only in the Tracking-ADMM local sub-problems are excluded from this comparison because they are algorithmic augmentation terms rather than components of the target CCP objective. Since intermediate distributed iterates need not yet satisfy the coupling constraint, their original objective values can lie on either side of the feasible centralized optimum, motivating the use of the absolute deviation. The remaining three quantities are those used by the finite distributed stopping rule in Section~\ref{sec:aggregation_mpc}: the maximum local coupling-feasibility residual, the maximum neighbour tracking disagreement, and the maximum penalty-scaled dual disagreement.

\begin{figure*}[t!]
\centering
\begin{subfigure}[t]{0.48\textwidth}
\centering
\includegraphics[width=\textwidth]{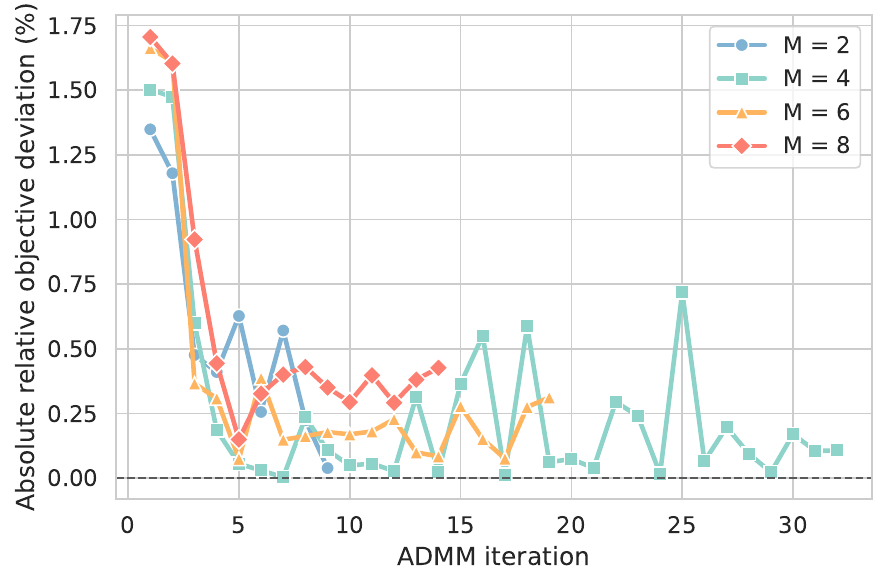}
\caption{Objective deviation.}
\end{subfigure}
\hfill
\begin{subfigure}[t]{0.48\textwidth}
\centering
\includegraphics[width=\textwidth]{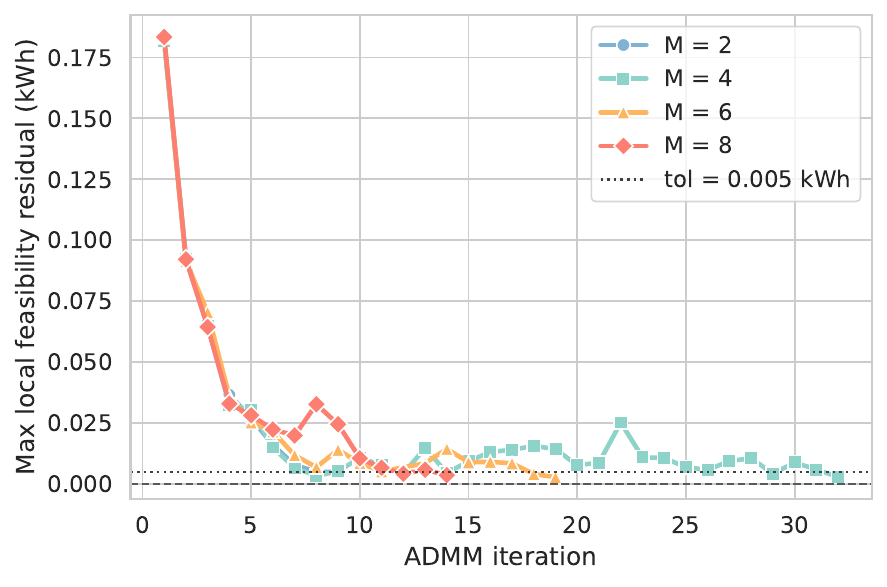}
\caption{Local coupling-feasibility residual.}
\end{subfigure}

\begin{subfigure}[t]{0.48\textwidth}
\centering
\includegraphics[width=\textwidth]{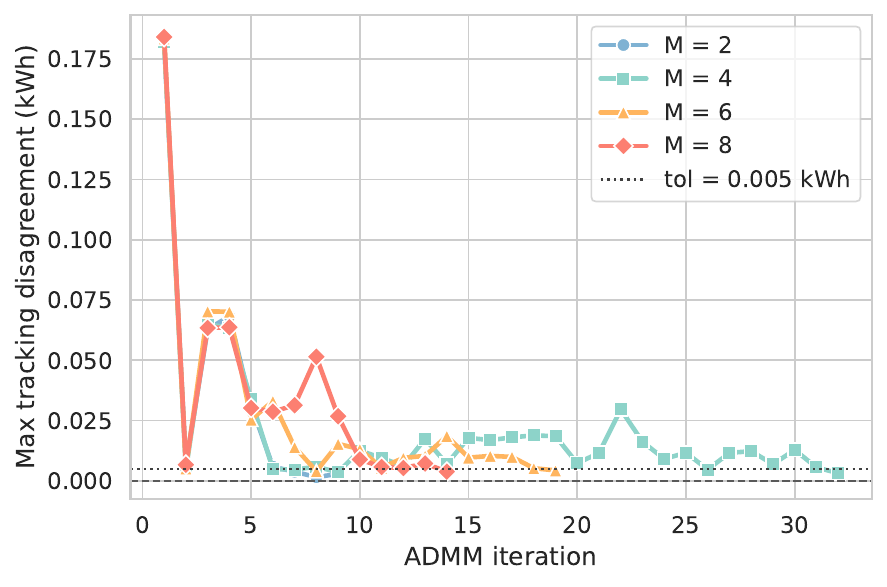}
\caption{Neighbour tracking disagreement.}
\end{subfigure}
\hfill
\begin{subfigure}[t]{0.48\textwidth}
\centering
\includegraphics[width=\textwidth]{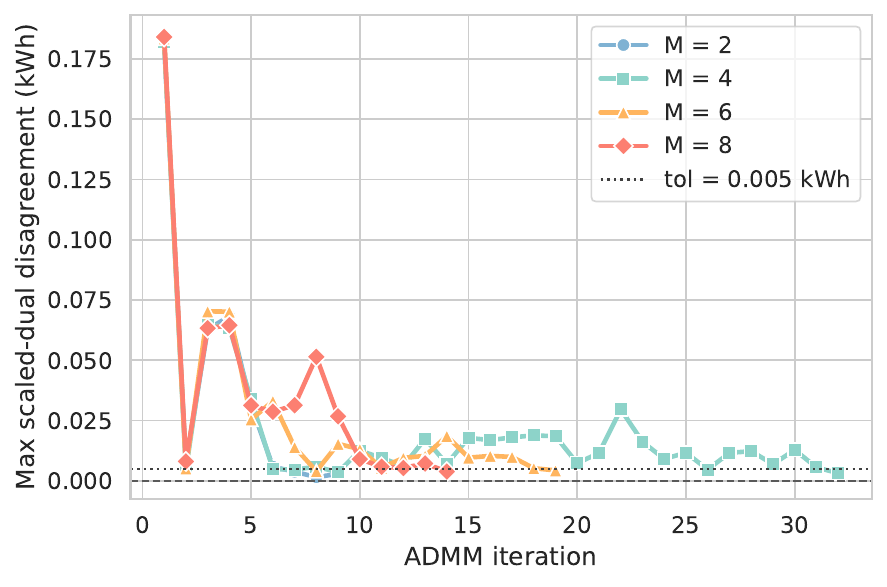}
\caption{Penalty-scaled dual disagreement.}
\end{subfigure}

\caption{Finite-iteration convergence of Tracking-ADMM at MPC update step 10 for symmetric aggregations with $N=2$, 4, 6, and 8. Panel (a) compares the original CCP objective evaluated at the distributed primal iterate with the corresponding centralized optimum. The dotted reference lines in (b)--(d) indicate the stopping tolerances used by the distributed implementation.}
\label{fig:convergence}
\end{figure*}

Figure~\ref{fig:convergence}(a) shows that the objective deviation decreases rapidly from its initial value for all four aggregation sizes and reaches the sub-percent range within the first few Tracking-ADMM iterations. The subsequent trajectories exhibit non-monotonic fluctuations rather than converging monotonically toward zero. This behaviour does not contradict convergence: the plotted quantity is the original CCP objective rather than the augmented objective minimized within each local Tracking-ADMM sub-problem, and intermediate primal iterates need not yet satisfy the aggregation coupling constraint. Objective proximity to the centralized optimum must therefore be interpreted together with the feasibility and consensus measures in Fig.~\ref{fig:convergence}(b)--(d).

Figure~\ref{fig:convergence}(b)--(d) shows the evolution of the quantities used by the finite distributed stopping rule. The local coupling-feasibility residual, neighbour tracking disagreement, and penalty-scaled dual disagreement all decrease substantially from their initial values and ultimately reach their prescribed tolerances. The number of iterations required at this particular MPC update is not monotonic in aggregation size, reflecting the dependence of Tracking-ADMM convergence speed on the specific optimization instance rather than on $N$ alone. Nevertheless, all four cases attain the distributed stopping conditions within the prescribed iteration budget. Importantly, termination depends only on these locally available and neighbour-based quantities; the centralized objective in Fig.~\ref{fig:convergence}(a) is used solely as an offline benchmark for assessing the quality of the distributed solution.

Overall, the results connect the asymptotic Tracking-ADMM guarantees in Section~\ref{sec:methodology} with the finite-iteration implementation in Section~\ref{sec:aggregation_mpc}. Across all tested aggregation sizes, the distributed iterates approach the centralized objective while the local feasibility and neighbour-consensus measures are driven to their prescribed stopping tolerances. The finitely terminated solutions are therefore not claimed to reproduce the centralized optimizer exactly; rather, the experiment verifies that the practical distributed stopping mechanism yields closely coordinated solutions whose original CCP objective remains close to the centralized optimum.

\subsection{Value of Aggregation-Level Coordination}
\label{ssec:coordination_value}

The preceding experiments establish the computational advantage of the distributed implementation and its convergence toward the centralized solution. The remaining question is therefore whether aggregation-level coordination itself provides a meaningful economic benefit. To isolate this effect, the present comparison uses Individual MPC and CMPC only: Individual MPC disables internal energy exchange, whereas CMPC provides the centralized coordinated benchmark. DMPC is not repeated in this comparison because its agreement with the centralized solution has already been evaluated in Section~\ref{ssec:scalability_convergence}.

The aggregation size is fixed at $N=4$, the largest tested size for which the mean CMPC solution time remains within the 15-min control interval. Five consumer--prosumer compositions are considered, with $n_p=0,\ldots,4$ and $n_c=4-n_p$. The PV-generation multiplier of 2.5 used in the scalability study is retained as the default condition, and a multiplier of 4.0 is additionally considered as a higher-PV sensitivity case. The multiplier itself is not the quantity expected to determine coordination value; the relevant quantity is the aggregation-level PV surplus available for internal redistribution, together with the consumer demand available to absorb that surplus.

Before enabling internal energy exchange, a separate no-aggregation experiment is used to establish the magnitude of closed-loop numerical variation. Internal trading is disabled in both Individual MPC and CMPC for $N=2$ with $(n_c,n_p)=(2,0),(1,1)$, and $(0,2)$. Under this restriction, the two controller formulations are mathematically equivalent: the centralized problem separates into the same independent local MPC problems solved by Individual MPC. Any remaining difference between their numerical solutions therefore cannot be attributed to aggregation-level coordination.

\begin{table}[t!]
\centering
\caption{No-aggregation numerical-equivalence check for $N=2$. Internal energy exchange is disabled in both controllers.}
\label{tab:noagg_check}
\small
\begin{tabular}{ccccc}
\toprule
$\boldsymbol{n_c}$ & $\boldsymbol{n_p}$ & \textbf{Individual (\texteuro)} & \textbf{CMPC (\texteuro)} & \textbf{Difference (\%)}\\
\midrule
2 & 0 & 27.821 & 27.906 & +0.31\\
1 & 1 & 21.774 & 21.744 & -0.14\\
0 & 2 & 15.843 & 15.976 & +0.84\\
\bottomrule
\end{tabular}
\end{table}

The reported percentage difference is calculated relative to the Individual-MPC electricity bill, with positive values indicating a higher bill under CMPC and negative values indicating a lower bill.

Despite this mathematical equivalence, Table~\ref{tab:noagg_check} shows small non-zero differences in the resulting full-day bills. The remaining differences arise because the mathematically equivalent problems are presented to IPOPT in different numerical forms and are solved only to finite tolerances. Individual MPC presents each building problem separately to IPOPT, whereas CMPC presents the mathematically equivalent collection as a joint block-structured nonlinear program. At MPC update step 10, before any controller-induced state divergence has occurred, both controllers use identical building histories and problem data and therefore solve mathematically equivalent no-trading problems; finite IPOPT tolerances can nevertheless lead to slightly different accepted solutions and first control actions. Once different actions are applied, the subsequent EnergyPlus states and prediction histories diverge, so later MPC problems are no longer identical and the initial numerical differences can accumulate over the closed-loop trajectory. The IPOPT tolerance of $10^{-4}$ used in the final experiments reflects the numerical-accuracy and computation-time trade-off identified during solver tuning; further tightening substantially increased solution time and was therefore not adopted in the reported experiments.

\begin{table*}[t!]
\centering
\caption{Economic value of centralized aggregation-level coordination for $N=4$ under different consumer--prosumer compositions and PV-surplus conditions. The baseline PV surplus is evaluated from the corresponding Individual-MPC trajectory. Positive savings indicate a lower electricity bill under CMPC than under Individual MPC.}
\label{tab:coordination_value}
\small
\begin{tabular}{cc c ccc}
\toprule
$\boldsymbol{n_c}$ & $\boldsymbol{n_p}$ & \textbf{Baseline PV surplus (kWh)} & \textbf{Individual (\texteuro)} & \textbf{CMPC (\texteuro)} & \textbf{Saving (\%)}\\
\midrule
\multicolumn{6}{l}{\textit{PV-generation multiplier = 2.5}}\\
4 & 0 & 0.00 & 55.50 & 55.31 & 0.34\\
3 & 1 & 7.32 & 49.27 & 49.18 & 0.18\\
2 & 2 & 19.25 & 43.38 & 42.75 & 1.45\\
1 & 3 & 28.44 & 37.33 & 36.30 & 2.76\\
0 & 4 & 37.13 & 31.40 & 31.32 & 0.25\\
\midrule
\multicolumn{6}{l}{\textit{PV-generation multiplier = 4.0}}\\
4 & 0 & 0.00 & 55.50 & 55.31 & 0.34\\
3 & 1 & 18.82 & 46.84 & 45.23 & 3.45\\
2 & 2 & 44.22 & 38.29 & 35.60 & 7.03\\
1 & 3 & 63.90 & 29.99 & 27.12 & 9.58\\
0 & 4 & 86.89 & 21.51 & 22.09 & -2.70\\
\bottomrule
\end{tabular}
\end{table*}

Under the default PV setting, the three mixed aggregations contain only a modest amount of redistributable surplus. The CMPC saving increases as more PV surplus becomes available, but some of these savings remain of the same order as the numerical closed-loop differences observed in the no-aggregation check. Consequently, small full-day bill differences under low-surplus conditions do not provide strong evidence of coordination value on their own.

The higher-PV case produces a substantially clearer result. Under the higher-PV setting, all three mixed compositions show a clear reduction in the centralized electricity bill relative to Individual MPC, with the coordination benefit becoming more pronounced as the available surplus increases. These improvements are clearly larger than the small numerical differences observed when internal trading is disabled, indicating that aggregation-level coordination becomes economically distinguishable once sufficient surplus energy is available for redistribution. 

PV surplus alone is nevertheless insufficient to create coordination value. When $n_p=0$, no prosumer surplus exists and internal trading has no energy source. Conversely, when $n_c=0$, the aggregation can contain substantial PV surplus but has no consumer demand to absorb internally redistributed energy. In both endpoint cases the internal trading opportunity therefore vanishes, and the remaining bill differences are attributable to the numerical closed-loop effects identified above. The economically relevant condition is thus the simultaneous presence of redistributable prosumer surplus and consumer demand capable of absorbing it.

The coordination-value experiment therefore shows that internal aggregation is economically beneficial when a meaningful surplus--demand matching opportunity exists; when that opportunity is weak or absent, the resulting economic difference can be comparable with the numerical variation of the closed-loop optimization.

\section{Conclusion and Future Work}
\label{sec:conclusion}

An MPC-aware feature-selection methodology was developed to evaluate candidate variables according to their future availability and recursive prediction requirements over the MPC horizon, with future-unknown selected features incorporated into the recursive prediction set so that their propagated errors are explicitly assessed. The selected feature sets are then used to train the previously developed IC-EoT, whose input-convex structure preserves the convexity of the resulting local MPC problems. The aggregation-level energy-balance constraint then gives rise to a convex constraint-coupled optimization problem, which is solved through Tracking-ADMM with bounded-neighbour communication and fully distributed finite-iteration stopping.

The experimental results support the framework from both modelling and optimization perspectives. Across both building testbeds, the proposed feature sets achieved the lowest horizon-mean weighted primary-target MSE and the lowest end-of-horizon rollout error, confirming the benefit of evaluating candidate variables under their actual recursive MPC deployment roles. In the multi-building experiments, centralized MPC became incompatible with the 15-min control interval at relatively modest aggregation sizes as its solution time increased sharply. In contrast, the DMPC solution time increased only mildly throughout the evaluated range and remained compatible with the 15-min control interval. This increase was accompanied primarily by a gradual rise in the number of Tracking-ADMM iterations required for distributed coordination at larger aggregation sizes. Where centralized references were available, finite-iteration DMPC approached the CMPC objective while satisfying the prescribed distributed stopping tolerances, and its closed-loop performance remained closely comparable with CMPC.

The coordination study further shows that, for a fixed aggregation size, the economic benefit of coordination depends on the amount of redistributable PV surplus and the availability of consumer demand to absorb it; the presence of prosumers alone does not guarantee a measurable bill reduction. When the available surplus is limited, the achievable bill reduction can be of the same order as the closed-loop numerical variation associated with finite solver accuracy, so a small coordination benefit can be masked in the realized bill. As the available prosumer surplus increases, a clear economic benefit emerges when sufficient consumer demand is available to absorb it. This interpretation is also consistent with the boundary cases: the all-consumer aggregation has no prosumer surplus available for internal redistribution, while the all-prosumer aggregation contains no consumer agents to receive internally traded energy under the market structure considered here.

Several limitations define the scope of these conclusions and motivate further work. First, the framework has been validated in high-fidelity EnergyPlus--Energym co-simulation rather than on physical buildings, and the aggregation heterogeneity considered here arises primarily from DER ownership and market roles while the agents share the same office-building archetype. Second, the online Tracking-ADMM implementation necessarily terminates after a finite number of iterations, whereas its theoretical convergence guarantee is asymptotic. The present controller therefore uses prescribed local-feasibility and neighbour-disagreement tolerances to determine distributed termination, and the experiments show that more iterations are generally required to satisfy these criteria as the aggregation size increases. Third, prediction uncertainty and model--plant mismatch are not explicitly accounted for within the MPC formulation. Future work will extend the framework to more diverse building and DER portfolios, investigate accelerated or adaptive Tracking-ADMM updates together with stopping criteria that provide explicit finite-iteration bounds on coupling feasibility and suboptimality, and develop uncertainty-aware distributed MPC formulations. Finally, hardware-in-the-loop and physical-building experiments would provide the next step toward validating the framework beyond simulation.

\section*{CRediT authorship contribution statement}

Kaipeng Xu: Conceptualization, Methodology, Software, Validation, Formal analysis, Investigation, Data curation, Visualization, Writing -- original draft, Writing -- review \& editing, Project administration. Zhuo Zhi: Software, Writing -- review \& editing. Keyue Jiang: Writing -- review \& editing.

\section*{Declaration of competing interest}

The authors declare that they have no known competing financial interests or personal relationships that could have appeared to influence the work reported in this paper.

\section*{Data availability}

Data and code supporting the findings of this study will be made available on reasonable request.

\section*{Declaration of generative AI and AI-assisted technologies in the manuscript preparation process}

During the preparation of this work, the authors used OpenAI ChatGPT and Google Gemini for manuscript organization, language refinement, writing support, and assistance with code development and debugging. The authors reviewed and edited the output as needed and take full responsibility for the content of the published article.

\bibliographystyle{elsarticle-num}
\bibliography{literature}

\end{document}